\documentclass[12pt]{article}
\usepackage{a4wide}

\usepackage[dvipdfmx,hiresbb]{graphicx} 
\usepackage{mediabb}                    

\usepackage{amsmath,amssymb}
\usepackage{color}
\usepackage[bookmarks,bookmarksnumbered]{hyperref}
\usepackage{cellspace}
\usepackage{tikz}
\usepackage{mathtools}
\usepackage{cite}
\usetikzlibrary{fadings}
\usetikzlibrary{patterns}
\usetikzlibrary{shadows.blur}
\usetikzlibrary{shapes}
\usepackage{ascmac}

\newcommand{\aln}[1]{\begin{align}#1\end{align}}
\newcommand{\nn}{\nonumber\\}

\begin{document}
\title{\vspace{-3cm}
\vbox{
\baselineskip 14pt
\hfill \hbox{\normalsize 
}}  \vskip 1cm
\bf \Large Classical Continuum Limit of the String Field Theory Dual to  Lattice Gauge Theory
\vskip 0.5cm
}
\author{
Kiyoharu Kawana\thanks{E-mail: \tt kkiyoharu@kias.re.kr}
\bigskip\\
\normalsize
\it 
School of Physics, Korean Institute for Advanced Study, Seoul 02455, Korea
\smallskip
}
\date{\today}

\maketitle   
\begin{abstract} 
 We discuss the classical continuum limit of the string field theory dual to  the $\mathrm{SU}(N)$ lattice gauge theory and investigate various fundamental phenomena in the continuum theory at the mean-field level. 
Our construction of the continuum theory is based on the concept of {\it area derivative}, which can be regarded as a generalization of the ordinary derivative $\partial/\partial x^\mu$ to operators acting on functional fields $\phi[C]$ on the loop space.  
The resultant continuum theory has a $\mathbb{Z}_N^{}$ $1$-form global symmetry, which originates in the $\mathbb{Z}_N^{}$ center symmetry in the gauge theory.  
We find that the confined and deconfined phases of the gauge theory are identified by the unbroken and broken phases of the $\mathbb{Z}_N^{}$ symmetry respectively by showing the Area/Perimeter law of the classical solution.       
In the broken phase, the low-energy effective theory is described by a $\mathrm{BF}$-type topological field theory and has a emergent $\mathbb{Z}_N^{}$ $(D-2)$-form global symmetry. 
The existence of the emergent symmetry is deeply related to $(D-2)$-dimensional topological configurations (i.e. center vortex for $D=4$), and we explicitly construct such a topological defect in the continuum theory.  
Finally, we also comment on the upper and lower critical dimensions of the gauge theory/string field theory.

\end{abstract} 

\setcounter{page}{1} 

\newpage  

\tableofcontents   

\newpage  

\section{Introduction}\label{Sec:intro}

In this paper, we discuss the classical continuum limit of the string field  theory dual to the $\mathrm{SU}(N)$ lattice gauge theory and perform a mean-field analysis of the continuum theory, emphasizing the role of the center group $\mathbb{Z}_N^{}$ as an $1$-form global symmetry. 
The dual string field theory was originally formulated in Refs.~\cite{Yoneya:1980bw,Banks:1980sq} by means of the gaussian integral transformation (i.e. Hubbard-Stratonovich transformation), with the expectation that it could be useful to discuss the critical behaviors of (lattice) gauge theories, similar to Landau theory for spin models.  
Although general aspects such as symmetries, topological excitations (defects), and large-$N$ behaviours were addressed in Ref.~\cite{Yoneya:1980bw}, the functional nature of the theory poses difficulties for practical calculations even at the mean-field level. 
Furthermore, understanding the continuum limit and corresponding continuum theory also remains as a critical issue in this approach.

On the other hand, there have been tremendous progresses on the fundamental natures of higher-dimensional objects in spacetime in the last half century.  
One of the developments is the generalization of global symmetry from  particles to higher-dimensional objects, referred to as ``higher-form global symmetry"~\cite{Gaiotto:2014kfa,Kapustin:2005py,Pantev:2005zs,Nussinov:2009zz,Banks:2010zn,Kapustin:2013uxa,Aharony:2013hda,Kapustin:2014gua,Gaiotto:2017yup,
McGreevy:2022oyu,Brennan:2023mmt,Bhardwaj:2023kri,Luo:2023ive,Gomes:2023ahz,Shao:2023gho}. 
The awareness of these generalized symmetries has broadened our understanding of quantum phases of matters and motivated us to explore new paradigms    beyond the conventional Landau theory for particles, i.e. $0$-form global symmetries.   

Another advancement relevant to this paper is the generalization of the ordinary derivative $\partial/\partial x^\mu$ to operators acting on functionals of higher-dimensional (closed) objects, which is referred to as the ``{\it area derivative}" in the literatures~\cite{Migdal:1983qrz,Makeenko:1980vm,Polyakov:1980ca,Kawai:1980qq,Rey:1989ti,Iqbal:2021rkn,Hidaka:2023gwh,Kawana:2024fsn}. 
In general, the area derivative describes the variation of a functional $\phi[C]$ of a closed higher-dimensional object $C$ with respect to a small change of the object $C+\delta C$.~(See Fig.~\ref{fig:area-derivative} for an intuitive image in the case of closed string.)
In the pioneering work~\cite{Iqbal:2021rkn} and the subsequent studies~\cite{Hidaka:2023gwh,Kawana:2024fsn}, a novel field theory for  general $p$-form global symmetries was investigated by employing the   
area derivative and it was found that various fundamental phenomena, such as the Area/Perimeter laws, Nambu-Goldstone theorem, topological defects and orders, etc. are elegantly and systematically explained, analogous to the conventional Landau theory for $0$-form symmetries.  

In light of these developments, we examine the classical continuum limit of the dual string field theory of the $\mathrm{SU}(N)$ lattice gauge theory and conduct a detailed analysis of its mean-field dynamics. 
%
We derive a continuum string field theory Eq.~(\ref{continuum action}) or (\ref{Lorentzian action}) at the quadratic order of the area derivative,
and find that its kinetic term is expressed by the d'Alembert operator constructed from the area derivatives. 
%
%
In this formulation, the $\mathbb{Z}_N^{}$ center symmetry of the original gauge theory is manifestly realized as a $\mathbb{Z}_N^{}$ $1$-form global symmetry due to an unique property of the area derivative, i.e. Eq.~(\ref{property of area derivative}). 
After the discussion of the classical continuum limit, we perform the mean-field analysis of the continuum string field theory. 
Following the similar discussions to Refs.~\cite{Iqbal:2021rkn,Hidaka:2023gwh,Kawana:2024fsn}, we see that the classical solution $\phi_{\rm cl}^{}[C]$ exhibits the Area (Perimeter) law in the unbroken (broken) phase of the $\mathbb{Z}_N^{}$ $1$-form global symmetry.  
Since the trace of the string field $\mathrm{Tr}(\phi[C])$ corresponds to the Wilson loop operator $W[C]$ in the gauge theory, these results imply that the confinement/deconfinement transition is now described in terms of the spontaneous breaking of the $\mathbb{Z}_N^{}$ $1$-form global symmetry in the dual sting field theory.  

In addition to these mean-field analyses, we also provide an explicit construction of a topological defect in the broken phase, which corresponds to the  center vortex in the gauge theory~\cite{Greensite:2003bk,Yoneya:1978dt}. 
Such a topological defect is a codimension $2$ object in spacetime and can be regarded as the symmetry operator (Gukov-Witten operator) of the $\mathbb{Z}_N^{}$ $1$-form global symmetry. 
The presence of codimension $2$ topological defects indicates the emergence of a $\mathbb{Z}_N^{}$ $(D-2)$-form global symmetry ($D$=spacetime dimension), and it is explicitly shown that the low-energy effective theory in the broken phase is a $\mathrm{BF}$-type topological field theory coupled to the topological defects. 
This theory has $\mathbb{Z}_N^{}$ $1$- and $(D-2)$-form global symmetries,  and can exhibit topological order when the spatial manifold has a nontrivial topology so that the symmetry operators can wrap around.     
Finally, we also comment on the critical dimensions of the gauge/string field theory.

\

The organization of this paper is as follows. 
In Sec.~\ref{sec2}, we review the lattice gauge theory and dual string field formulation.   
Compared to the original work~\cite{Yoneya:1980bw}, we refine various quantities, including  the definition of the kinetic operator (and weight functional $w[C]$) to be appropriate for the classical continuum limit.     
In Sec.~\ref{sec3}, we discuss the classical continuum limit of the dual string field theory and obtain the continuum string field theory.       
We then perform a mean-field analysis of the continuum theory and construct a  topological defect explicitly in Section~\ref{sec4}. 
Additionally we comment on the critical dimensions of the present theory. 
Section~\ref{sec5} is devoted to summary and discussion. 
The Appendices provide supplementary details of the discussion.  
%


\section{Dual String Field Theory of Lattice Gauge Theory}\label{sec2}

We review the derivation of the dual string field theory of the $\mathrm{SU}(N)$ lattice gauge theory~\cite{Yoneya:1980bw,Banks:1980sq}. 
The flow of the derivation is essentially the same as the original argument,  but we refine the definition of the kinetic operator $\hat{H}$ of the string field to be appropriate for the continuum limit later.  
In addition, we present a concrete example of the weight functional $w[C]$  that is suitable for the mean-field analysis.   

The reader who is familiar with the lattice gauge theory and dual formulation can skip this section and move to the next section.

\subsection{Formulations}

\begin{figure}
    \centering
    \includegraphics[scale=0.3]{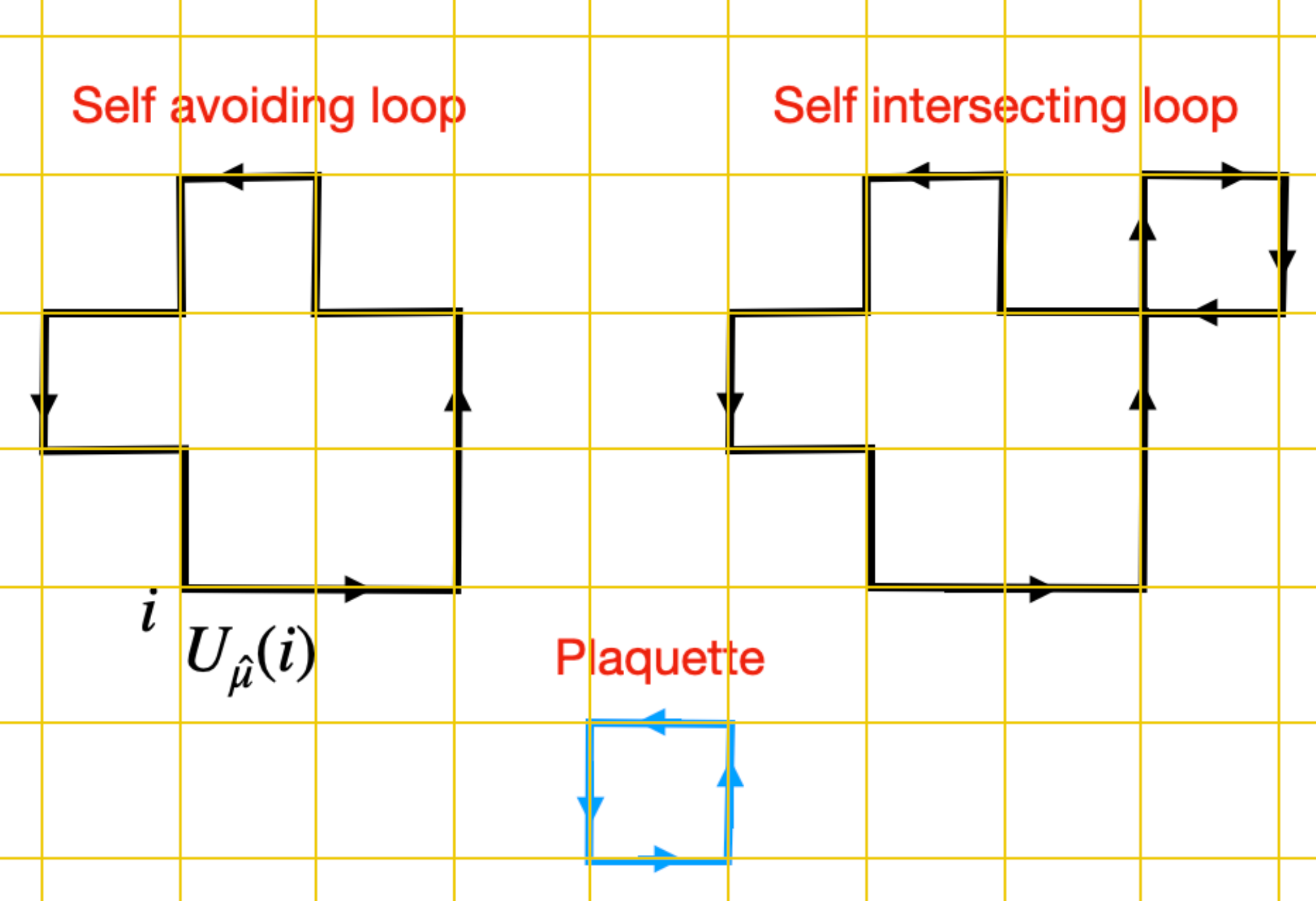}
\caption{
 A self-avoiding loop (left) and self-intersecting loop (right). Here arrows represent the orientation of a loop and a plaquette is represented by blue in general.}
 \label{fig:lattice}
\end{figure}

We consider a $D$-dimensional Euclidean cubic lattice $\Lambda_D^{}$ with a lattice spacing $a$ and take $G=\mathrm{SU}(N)$ as a gauge group.  
In the following, we focus on {\it self-avoiding} loop $C$ with an orientation such that its path never revisits previously visited sites  
as shown in Fig.~\ref{fig:lattice}. 
The set of all self-avoiding loops is represented by $\Gamma$. 
For a given loop $C\in \Gamma$, the total number of sites along the loop is denoted by $|C|$. 
The loop with the opposite orientation of $C$ is represented as $C^{-1}\in \Gamma$. 
A minimum square loop $P$ is called the {\it plaquette} as usual. 
In particular, the orientation of a plaquette $P$ is defined as {\it positive~(negative)} when it is counterclockwise (clockwise).   
Moreover, a general $1$-form field $B_1^{}$ is written as 
\aln{
B_1^{}=B_\mu^{}(i)a\hat{e}^\mu\rightarrow B_\mu^{}(x)dx^\mu\quad (\text{in the continuum limit})~, 
}    
where $i\in \Lambda_D^{}$ denotes a site and $\hat{e}^\mu~(\mu=1,2,\cdots ,D)$ is the unit basis.  
In this notation, the exterior derivative is expressed as  
\aln{
dB_1^{}&\coloneqq \frac{B_\nu^{}(i+a\hat{\mu})-B_\nu^{}(i)}{a}a\hat{e}^\mu\wedge a\hat{e}^\nu
\nn
&\rightarrow \frac{1}{2}(dB_1^{})_{\mu\nu}^{}dx^\mu\wedge dx^\nu\coloneq 
\partial_\mu^{}B_\nu^{}(x)dx^\mu\wedge dx^\nu\quad (\text{in the continuum limit})~,
}
where $\hat{\mu}=(0,\cdots,0,1,0,\cdots,0)$ is the unit vector in the $\mu$ direction. 
We represent its component as $(\hat{\mu})^\nu=\delta^\nu_\mu$ when necessary. 
For a given lattice site $i\in \Lambda_D^{}$, we introduce a link variable by 
\aln{
U_{\hat{\mu}}^{}(i)\coloneqq \exp\left(i a A_{\mu^{}}(i)\right)~,
}
where $A_{\mu}^{}(i)$ is an element of the Lie algebra of $\mathrm{SU}(N)$.   
We define the inverse link variable by
\aln{U_{-\hat{\mu}}(i+\hat{\mu})\coloneqq U^\dagger_{\hat{\mu}}(i)~.
\label{inverse link}
}
The gauge transformation of a link variable is 
\aln{
U_{\hat{\mu}}(i)\quad \rightarrow \quad g_{}^{}(i)U_{\hat{\mu}}(i)g^\dagger(i+\hat{\mu})~,\quad g(i)\in \mathrm{SU}(N)~. 
\label{gauge transformation}
}
For a loop $C\in \Gamma$ consisting of a sequence of sites $(i=i_1^{},\cdots,i_{|C|}^{})$, we define the path-ordered product by  
\aln{
U[C;i]&\coloneqq \prod_{l=1}^{|C|}U_{\hat{\mu}_l^{}}^{}(i_l^{})
\label{path-ordered product}
\\
&=\mathrm{P}\exp\left(i\oint_{C} A_1^{}\right)\quad (\text{in the continuum limit})~,
} 
where 
$a\hat{\mu}_l^{}=i_{l+1}^{}-i_{l}^{}$ and $\mathrm{P}$ here denotes the path ordering.    
For a plaquette $P$ placed in the $\mu$-$\nu$ plane, we also express the corresponding path-ordered product as  
\aln{
U_{\mu\nu}^{}(i)\coloneqq U^{}[P;i]&=U_{\hat{\mu}}^{}(i)U_{\hat{\nu}}^{}(i+\hat{\mu})U_{-\hat{\mu}}^{}(i+\hat{\mu}+\hat{\nu})U_{-\hat{\nu}}^{}(i+\hat{\nu})
\\
&=U_{\hat{\mu}}^{}(i)U_{\hat{\nu}}^{}(i+\hat{\mu})U^\dagger_{\hat{\mu}}(i+\hat{\nu})U^\dagger_{\hat{\nu}}(i)~.
}
A Wilson-loop operator is defined by taking the trace of Eq.~(\ref{path-ordered product}) as 
\aln{
W[C]={\rm Tr}\left(U[C;i]\right)~,
}
which is independent of the initial site $i$. 
The partition function of the lattice gauge theory is defined by
\aln{
Z[J]&=\int [dU]\exp\left(-S_\mathrm{E}^{}\right)~,\label{lattice partition function}
\\
S_\mathrm{E}^{}&=\beta\sum_{i\in \Lambda^{}}\sum_{\mu>\nu}^D \left[1-\frac{1}{2N}\mathrm{Tr}\left(U_{\mu\nu}(i)+U_{\mu\nu}^\dagger (i)
\right)\right]
-\sum_{C^{}\in \Gamma}^{}J[C]^*W[C]+{\rm h.c.}~
\\
&=\beta\sum_{P:\text{positive}}\left[1-\frac{1}{2N}\left(W[P]+W^\dagger[P]\right)\right]-\sum_{C^{}\in \Gamma}J[C]^*W[C]+{\rm h.c.}~,
\label{lattice action}
}
where the summation $\sum_{P:\text{positive}}$ implies the sum of all  positive oriented plaquettes, $J[C]$ is an external source for the Wilson loop, and 
\aln{[dU]\coloneq \prod_{i\in \Lambda_D^{}}\prod_{\hat{\mu}=1}^D dU_{\hat{\mu}}(i)~, 
}
 is the Haar measure of $\mathrm{SU}(N)$ satisfying 
\aln{
dU_{\hat{\mu}}^{}(i)=d(VU_{\hat{\mu}}^{}(i))=d(U_{\hat{\mu}}^{}(i)V)
}
for $^\forall V\in \mathrm{SU}(N)$. 
One can see that this measure is invariant under the gauge transformation (\ref{gauge transformation}) because of this property.    
Besides, 
\aln{\beta=\frac{2N}{g^2}
} 
is the inverse gauge coupling. 

\begin{figure}
    \centering
    \includegraphics[scale=0.2]{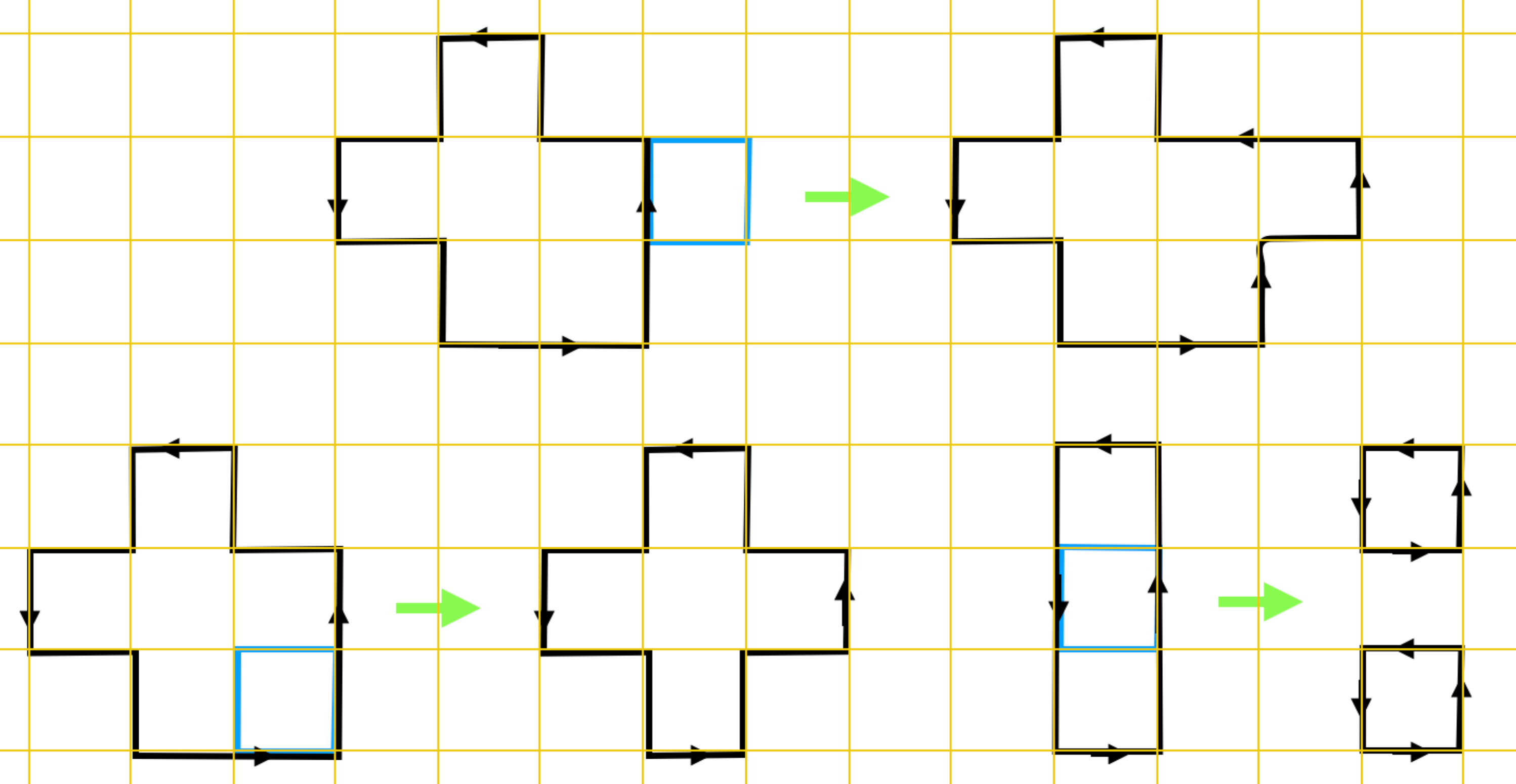}
    \caption{
    Combination of a self-avoiding coop $C$ (black) and plaquette $P$ (blue). 
    The upper and lower-left ones correspond to $C+P\in \Gamma$, while the lower-right corresponds to $C+P\notin \Gamma$. 
    }
    \label{fig:sum}
\end{figure}

For a self-avoiding loop $C\in \Gamma$ and plaquette $P$, we define a combined loop $C+P$ by erasing the common links between $C$ and $P$ and attaching the links of $P$ which are not common with $C$. 
See Fig.~\ref{fig:sum} for a few examples.  
Note that, for a common link, we can attach either positive or negative oriented plaquette.  
An important property is 
\aln{
{\rm Tr}(U^\dagger[C;i]U[C+P;i])=W[P]\quad (\text{for } ^\forall i\in C)
\label{U property}
}
when $C+P\in \Gamma$. 
When $C+P\notin \Gamma$, Eq.~(\ref{U property}) does not necessarily hold as can be seen in the lower-right figure in Fig.~\ref{fig:sum}.

Now we consider a general $N\times N$ complex matrix functional $\phi[C]$ defined on the loop space $\Gamma$.   
We assume that its complex conjugate is related to the inverse loop by 
\aln{
\phi^\dagger [C]=\phi[C^{-1}]~.
}
%
%
We introduce a translation operator along a plaquette $P$ as~\cite{Yoneya:1980bw,Banks:1980sq} 
\aln{
\Pi_P^{}\phi[C]\coloneq \begin{cases}\phi[C+P] & \text{for } C+P\in \Gamma
\\
0 & \text{for } C+P\notin \Gamma 
\end{cases}~.
}
By using this translation operator, we can express the lattice action (\ref{lattice action}) as
\aln{
S_\mathrm{E}^{}&=\text{constant}-\frac{\beta}{2N}\sum_{C^{}\in \Gamma }\mathrm{Tr}\left(w[C]U^\dagger[C;i] (k_0^{}+\hat{H})w[C]^*U[C;i]
\right)-\sum_{C\in \Gamma}J[C]^*W[C]
+{\rm h.c.}~.
\label{quadratic action}
}
Here, $i\in C$ is an arbitrary reference site, and 
the operator $\hat{H}$ is defined by 
\aln{
\hat{H}\phi[C]\coloneqq \frac{1}{2(D-1)|C|}\sum_{P\in C}\Pi_P^{}\phi[C]~,
\label{derivative operator}
}
where the summation implies the sum over all different plaquettes which have common links with $C$. 
Besides, $k_0^{}$ is a sufficiently large positive constant to make $k_0^{}+\hat{H}$ positive definite.  
Introduction of $k_0^{}$ does not change any observables because it just corresponds to a constant shift of the action.   
Note also that the normalization of $\hat{H}$ is chosen so that $\hat{H}\rightarrow 1$ in the continuum limit $a\rightarrow 0$.   

Moreover, $w[C]$ is an arbitrary weight functional that is invariant under space(time) translation and satisfies 
%
\aln{
\frac{1}{2(D-1)}\sum_{C\in P}\frac{1}{|C|}w[C]w[C+P]^*=1~,
\label{w normalization}
}
%
where the summation means the sum over all different loops $C$ which have common links with a given plaquette $P$ such that both $C$ and $C+P$ are self-avoiding.   
A simplest choice of the weight functional is
\aln{w[C]= b\times W_{\rm abe}^{}[C]~,
\label{simple weight function}
}
where $ W_{\rm abe}^{}[C]$ is the Abelian Wilson loop constructed by a closed $1$-form 
\aln{A_1^{}=\Lambda_1^{}~,\quad d\Lambda_1^{}=0~,\quad \frac{1}{2\pi }\oint_C \Lambda_1^{}\in \mathbb{Z} ~,
} 
and $b$ is a normalization factor determined by Eq.~(\ref{w normalization}), i.e. 
\aln{
\frac{b^2}{2(D-1)}\sum_{C\in P}\frac{1}{|C|}e^{-i\int_P^{}\Lambda_1^{}}=\frac{b^2}{2(D-1)}\sum_{C\in P}\frac{1}{|C|}=1\quad  \therefore~b^2=\frac{2(D-1)}{\sum_{C\in P}1/|C|}~,
}
where we have used the Stokes theorem. 
An advantage of Eq.~(\ref{simple weight function}) is $w[C]^*w[C]=b^2=$constant, which leads to a simple expression of the potential in the dual string field theory. 
See Sec.~\ref{sec:potential} for more details.  
The equivalence between Eq.~(\ref{lattice action}) and (\ref{quadratic action}) can be checked as 
\aln{
\sum_{C\in \Gamma }\mathrm{Tr}\left(w[C]U^\dagger[C;i]) \hat{H}w[C]^*U[C;i]
\right)
&=\sum_{C\in \Gamma }\mathrm{Tr}\left(w[C]U^\dagger [C;i]\frac{1}{2(D-1)|C|}\sum_{P\in C}w[C+P]^*U[C+P;i]
\right)
\nn
&=\sum_{P}W[P]\frac{1}{2(D-1)}\sum_{C\in P}\frac{1}{|C|}w[C]w[C+P]^*
\nn
&=\sum_{P:\text{positive}}\mathrm{Tr}(W[P]+W^\dagger[P])~,
\label{proof}
}
where we have used Eq.~(\ref{U property}) (Eq.~(\ref{w normalization})) from the first (second) line to the second (third) line. 
This also confirms that Eq.~(\ref{quadratic action}) is independent of a reference site $i\in C$. 
Note also that $\hat{H}$ is self-adjoint as follows: 
\aln{
\sum_{C\in \Gamma}{\rm Tr}\left(\phi_1^\dagger[C]\hat{H}\phi_2^{}[C]\right)&=\sum_{C\in \Gamma}\sum_{P\in C}\mathrm{Tr}\left(\phi_1^\dagger[C]\phi_2^{}[C+P]\right)
\nn
&=\sum_{P}\sum_{C\in P}\mathrm{Tr}\left(\phi_1^\dagger[C]\phi_2^{}[C+P]\right)~.
\label{self adjointness}
}
%
Since $C'\coloneqq C+P$ is also a self-avoiding loop by the definition of $\Pi_P^{}$, Eq.~(\ref{self adjointness}) can be written as
\aln{
=\sum_P \sum_{C'\in P}\mathrm{Tr}\left(\phi_1^\dagger[C'+P^{-1}]\phi_2^{}[C']\right)=\sum_{C\in \Gamma}\mathrm{Tr}\left(\left\{\hat{H}\phi_1^\dagger[C]\right\}\phi_2^{}[C]\right)~,
}
which confirms $\hat{H}^\dagger =\hat{H}$. 

Since Eq.~(\ref{quadratic action}) is a quadratic form, we can perform the Hubbard-Stratonovich transformation as 
\aln{
&Z[J]\propto \int [dU]\int [d\phi]\exp\bigg\{-\frac{2N}{\beta}\sum_{C\in \Gamma}\mathrm{Tr}\left(\phi^\dagger[C](k_0^{}+\hat{H})^{-1}\phi[C]\right)~
\nn
&-\frac{2N^2}{\beta}\sum_{C\in \Gamma}\left(\frac{J[C]}{w[C]}\right)^*(k_0^{}+\hat{H})^{-1}\frac{J[C]}{w[C]}-\sum_{C\in \Gamma}\mathrm{Tr}\left[\left(w[C]^*U[C;i]+(k_0^{}+\hat{H})^{-1}\frac{2NJ[C]}{\beta w[C]}
\right)^\dagger \phi[C]\right]+{\rm h.c.}\bigg\}
\\
=&\int [d\phi]\exp\left(-\frac{2N}{\beta}\sum_{C\in \Gamma}\mathrm{Tr}\left[\left(\phi[C]+\frac{J[C]}{w[C]}I_N^{}\right)^\dagger (k_0^{}+\hat{H})^{-1}\left(\phi[C]+\frac{J[C]}{w[C]}I_N^{}\right)\right]-V[\phi,\phi^\dagger]\right)~,
\label{loop functional theory}
} 
where 
\aln{
V[\phi,\phi^\dagger]=-\log\left[\int [dU]\exp\left(-\sum_{C\in \Gamma}w[C]\mathrm{Tr}\left(U^\dagger [C;i]\phi[C]\right)+{\rm h.c.}\right)\right]~
\label{psi potential}
}
is the potential of $\phi[C]$. 
We should note that this does not depend on the choice of a reference  site $i\in C$ as well.~(See Section~\ref{sec:potential} for more details.) 
One can see that the lattice gauge theory is now cast into a functional theory on the loop space, i.e. a string field theory. 
%

\subsection{Symmetries}
Let us discuss the symmetries of Eq.~(\ref{loop functional theory}) with $J[C]=0$.  

\

\noindent {\bf GAUGE SYMMETRY}\\
From Eq.~(\ref{psi potential}), one can see that the original gauge symmetry (\ref{gauge transformation}) is now represented by   
\aln{
U[C;i]\quad &\rightarrow \quad g(i)U[C;i]g^\dagger (i)~,
\\
\phi[C]\quad &\rightarrow \quad g(i)\phi[C]g^\dagger(i)~. 
}  
The kinetic term in Eq.~(\ref{loop functional theory}) is invariant under this transformation because $\hat{H}$ does not shift sites. 
%
%

\

\noindent {\bf $\mathbb{Z}_N^{}$ 1-FORM SYMMETRY}\\
In addition, Eq.~(\ref{loop functional theory}) has another global symmetry given by\footnote{In Ref.~\cite{Yoneya:1980bw}, this symmetry was regarded as ``local" symmetry because the group element $g[C]$ depends on loops $C$ in general.
} 
\aln{
\phi[C]&\quad \rightarrow \quad \phi'[C]=g[C]\phi[C]=\phi[C]g[C]~,
\label{1-form symmetry}
}
where $g[C]\in \mathbb{Z}_N^{}$ is an element of the centre $\mathbb{Z}_N^{}$ of $\mathrm{SU}(N)$.  
Note that $g[C]\in \mathbb{Z}_N^{}$ keeps the conjugation relation ${\phi'}^\dagger[C]=\phi'[C^{-1}]$. 
In order to see the invariance of the action under Eq.~(\ref{1-form symmetry}), we should first note that $g[C]$ can be regarded as a twisted  gauge transformation on $C$ in the following way: 
Let us explicitly represent $g[C]$ as  
\aln{
g[C]=\exp \left(\frac{2\pi i }{N}n\hat{T}\right)=e^{\frac{2\pi i}{N}n}I_N^{}~,\quad n\in \mathbb{Z}~,
\label{center transformation}
}
where $\hat{T}$ is the generator of the center group.\footnote{For example, we can take $\hat{T}={\rm diag}(1,1,\cdots,1,1-N)$.
}    
Then, we can introduce a local function $\theta(i)$ such that 
\aln{
g(j)=\exp\left(\frac{\theta(j)}{N}\hat{T}\right)~,\quad j\in \Lambda~,
}
with
\aln{
\theta(i_{|C|+1}^{}=i_1^{})=\theta(i_1^{})+2\pi n\quad \leftrightarrow\quad \oint_C d\theta\coloneqq a\sum_{k=1}^{|C|} \frac{\theta(i_{k+1}^{})-\theta(i_k^{})}{a}=2\pi n~,\quad n\in \mathbb{Z}~,
}
where $\{i_k^{}\}_{k=1}^{|C|}$ denotes a sequence of sites on $C$. 
Now one can see
\aln{g(i_1^{})\phi[C]g^\dagger(i_{|C|+1}^{})=\exp\left(\frac{i}{N}\hat{T}\oint_{C}d\theta \right)\phi[C]=g[C]\phi[C]~,
\label{twisted gauge transformation}
}
or equivalently 
\aln{
g[C]=W_{\rm abe}^{}[C]\bigg|_{A_1^{}=d\theta}^{}\times I_N^{}~. 
} 
This then leads to an commutative relation  
\aln{
g^\dagger [C]\hat{H}g[C]=\frac{1}{2(D-1)|C|}\sum_{P\in C^{}}g^\dagger [C]g[C+P]=\frac{1}{2(D-1)|C|}\sum_{P\in C} W_{\rm abe}^{}[P]\bigg|_{A_1^{}=d\theta}^{}\times I_N^{}=I_N^{}~
\label{abelian Poincare lemma}
}
where we have used the Poincare lemma.    
This commutative relation obviously implies the invariance of the kinetic term in Eq.~(\ref{loop functional theory}) under the $\mathbb{Z}_N^{}$ transformation (\ref{1-form symmetry}).
On the other hand, the invariance of the potential~(\ref{psi potential}) is much easier to check because the transformation~(\ref{1-form symmetry}) or (\ref{twisted gauge transformation}) can be absorbed into $U[C;i]$ as a (twisted) gauge transformation  without changing the path-integral measure. 
%
%
%

In modern physics, a symmetry given by the transformations such as  Eq.~(\ref{1-form symmetry}) is referred to as a (discrete) $1$-form global symmetry.  
Of course, the existence of this $1$-form symmetry in the dual string field theory has been anticipated from the beginning, as the original gauge theory possesses it as a center symmetry.    
%
Indeed one can see that Eq.~(\ref{twisted gauge transformation}) corresponds to a $\mathbb{Z}_N^{}$ transformation of the Wilson loop as  
\aln{
W[C]\quad \rightarrow \quad \exp\left(\frac{2\pi i}{N}n\right)W[C]~.
\label{Wilson loop transformation}
} 
What is the symmetry operator in this case ? 
%
It can be formally introduced by a $(D-2)$-dimensional (topological) operator $U_{\rm GW}^{}[C_{D-2}^{}]$ such that it provides the $\mathbb{Z}_N^{}$ transformation (\ref{Wilson loop transformation}) when it is linked with the Wilson loop:
\aln{
U_{\rm GW}^{}[C_{D-2}^{}]W[C]=e^{i\frac{2\pi}{N}\mathrm{Link}[C_{D-2}^{},C]}W[C]U_{\rm GW}^{}[C_{D-2}^{}]~,
 }
 where $\mathrm{Link}[C_{D-2}^{},C]$ is the linking number between $C_{D-2}^{}$ and $C$. 
 In the lattice gauge theory (\ref{lattice action}), we can actually construct such a topological configuration as a local minimum of the action, known as the center vortex~\cite{Greensite:2003bk,Yoneya:1978dt}. 
See also the discussion in Sec.~\ref{defect} and Appendix~\ref{center vortex} for the center vortex solution in the lattice gauge theory.  
%


\subsection{Structure of the potential}\label{sec:potential}

\begin{figure}
    \centering
    \includegraphics[scale=0.35]{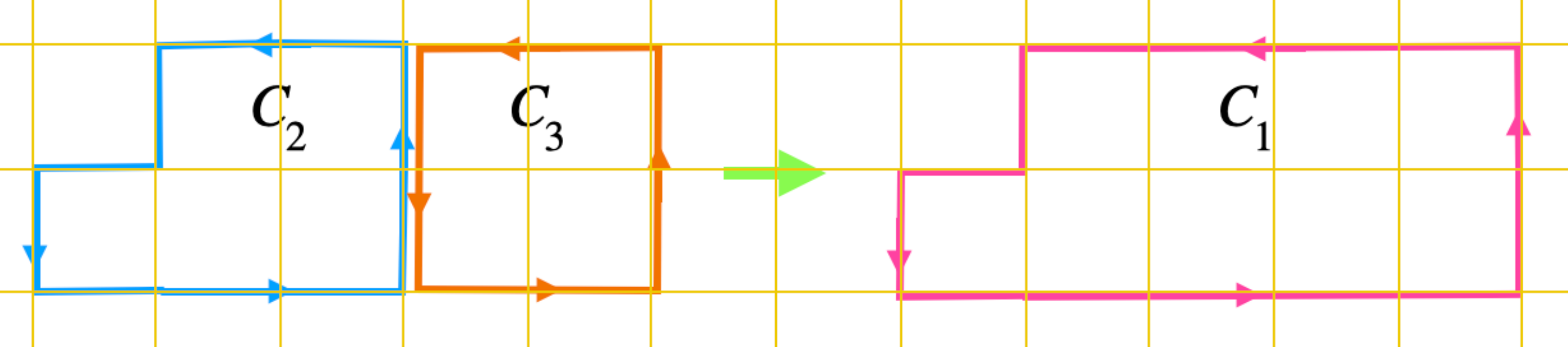}
    \caption{Sum of two loops with same orientation. 
    }
    \label{fig:interaction}
\end{figure}

The existence of the $\mathbb{Z}_N^{}$ $1$-form global symmetry constraints  the form of the potential (\ref{psi potential}), and it can be generally expanded as 
\aln{
V[\phi,\phi^\dagger]=&\sum_{C\in \Gamma}\left[\mu_2^{}[C]\mathrm{Tr}(\phi^\dagger[C]\phi[C])+\mu_2^{'}[C]\mathrm{Tr}(\phi^\dagger[C])\mathrm{Tr}(\phi[C])\right]
\nn
+\sum_{C_1^{}\in \Gamma}&\sum_{C_2^{}\in \Gamma}\sum_{C_3^{}\in \Gamma}\delta(C_1^{}-C_2^{}-C_3^{})\bigg[\mu_3^{}[\{C_i^{}\}]\mathrm{Tr}(\phi[C_1^{}]^\dagger \phi[C_2^{}]\phi[C_3^{}])+\mu_3^{'}[\{C_i^{}\}]\mathrm{Tr}(\phi^\dagger [C_1^{}]\phi[C_2^{}])\mathrm{Tr}(\phi[C_3^{}])
\nn
&+\mu_3^{''}[\{C_i^{}\}]\mathrm{Tr}(\phi^\dagger [C_1^{}])\mathrm{Tr}(\phi[C_2^{}])\mathrm{Tr}(\phi[C_3^{}])\bigg]+\text{h.c.}+{\cal O}(|\phi|^4)~,
\label{potential expansion}
}
where $\delta (C_1^{}+C_2^{}+C_3^{})$ is the delta function in the loop space  that restricts the loop configurations to $C_2^{}+C_3^{}=C_1^{}$ as shown in Fig.~\ref{fig:interaction}.   
%
%
The existence of the delta function ensures the invariance of the cubic terms under the $\mathbb{Z}_N^{}$ global $1$-form transformation as 
\aln{
 \delta(C_1^{}-C_2^{}-C_3^{}) g^\dagger[C_1^{}]g[C_2^{}]g[C_3^{}]&= \delta(C_1^{}-C_2^{}-C_3^{})\exp\left(\frac{2\pi i}{N}\hat{T}\oint_{-C_1^{}+C_2^{}+C_3^{}}d\theta 
\right)I_N^{}
\nn
&=\delta(C_1^{}-C_2^{}-C_3^{})I_N^{}~.
\label{cancellation of g}
}
The same cancelation mechanism applies to higher-order non-local interactions as well.   
%
%
We should note that the coupling constants in the expansion~(\ref{potential expansion}) are generally dependent on $C$ for a general weight functional $w[C]$.
%
%
This resembles the situation in ordinary field theory where coupling constants depend on spacetime points.    
However, the coupling constants in Eq.~(\ref{potential expansion}) are all independent of the center-of-mass position of $C_i^{}$, as long as $w[C]$ is a scalar under spacetime translation.  
In particular, one can see that these coupling constants become completely   independent of $C$ for the simplest weight functional (\ref{simple weight function}) as
\aln{
\begin{cases}\mu_2^{}[C]\propto w[C]^* w[C]=b^2
\\
\mu_3^{}[\{C_i^{}\}]\propto w[C_1^{}]^* w[C_2^{}]w[C_3^{}]=b^3
\end{cases}~,
}
where we have used the property that Eq.~(\ref{simple weight function}) also satisfies the same equation as~(\ref{cancellation of g}).   
%

Because of the definition (\ref{psi potential}), these coupling constants are not generally independent of each other and constrained by the functional equations such as 
\aln{
\frac{1}{w[C]^*w[C]}\frac{\delta}{\delta {(\phi[C])^{}}_{ik} }&\frac{\delta}{\delta {(\phi[C])^{\dagger}}_{kj}}\exp\left(-V[\phi,\phi^\dagger]\right)=\delta_{ij}\exp\left(-V[\phi,\phi^\dagger]\right)~,
\\
\delta(C_1^{}-C_2^{}-C_2^{})\frac{1}{w[C_2^{}]w[C_3^{}]}&\frac{\delta}{\delta {(\phi[C_2^{}])^{}}_{ik} }\frac{\delta}{\delta {(\phi[C_3^{}])}_{kj}}\exp\left(-V[\phi,\phi^\dagger]\right)
\nn
&=\delta(C_1^{}-C_2^{}-C_2^{})\frac{1}{w[C_1^{}]}\frac{\delta}{\delta {(\phi[C_1^{}])}_{ij}}\exp\left(-V[\phi,\phi^\dagger]\right)~,
\\
{\rm det}\left(\frac{1}{w[C]}\frac{\delta}{\delta (\phi[C])_{ij}^{}}\right)&\exp\left(-V[\phi,\phi^\dagger]\right)=\exp\left(-V[\phi,\phi^\dagger]\right)~, 
}
which can be directly checked from the definition (\ref{psi potential}).

Furthermore, we can also write down many other $\mathbb{Z}_N^{}$ invariant terms consisting of the product of $N$ elements of $\phi[C]$ such as 
\aln{
V[\phi,\phi^\dagger]\ni \sum_{C\in \Gamma}\bigg[&\alpha_N^{}[C]\mathrm{Tr}(\phi[C]^N)+\alpha_{N-1}^{}[C]\mathrm{Tr}(\phi[C]^{N-1})\mathrm{Tr}(\phi[C]^{})+\cdots+\alpha_0^{}[C](\mathrm{Tr}(\phi[C]^{}))^N
\nn
&+\beta_1^{}[C]\mathrm{det}(\phi[C])\bigg]+\cdots +{\rm h.c.}~
\label{determinant terms}
}
in general. 
%

\section{Classical Continuum Theory}\label{sec3}
Now, let us discuss the classical continuum limit of the dual string field theory~(\ref{loop functional theory}).  
In the continuum limit $a\rightarrow 0$, a self-avoiding loop $C\in \Gamma$ is represented by a set of embedding functions 
\aln{
X^\mu: S^1\quad \rightarrow \quad \mathbb{R}^D~,\quad \mu=1,2,\cdots,D
}
where $S^1$ is the $1$-dimensional sphere whose general intrinsic coordinate is represented by $\xi\in [0,2\pi)$.  
We will see that our continuum action (i.e. Eq.~(\ref{continuum action})) can be also defined on Lorentzian spacetimes by performing the Wick rotation.  
In general, we represent a $D$-dimensional Lorentzian spacetime with the metric $g_{\mu\nu}^{}(X)$ by $\Sigma_D^{}$, and employ the Minkowski metric signature as $(-,+,\cdots,+)$. 
Then, the induced metric on $C$ is given by 
\aln{
ds^2=g_{\mu\nu}^{}(X)dX^\mu dX^\nu=g_{\mu\nu}^{}(X(\xi))\frac{dX^\mu}{d\xi} \frac{dX^\nu}{d\xi}d\xi^2\coloneqq h(\xi)d\xi^2~,\quad h(\xi)>0~,
}
which leads to the continuum expression of the length of $C$ as
\aln{
L[C]\coloneqq a|C|\quad \rightarrow \quad \int_0^{2\pi}d\xi \sqrt{h(\xi)}~. 
}
This is invariant under the reparametrization $\xi~\rightarrow ~\xi'=f(\xi)$. 
In addition, the area of the minimal surface $M$ bounded by $C$, i.e. $\partial M=C$ is written as 
\aln{
{\rm Area}(C)=\int d^2\theta \sqrt{\gamma(\theta)}=\int_{M}E_2^{}~,
\label{minimal surface}
}
where
\aln{
E_2^{}= \frac{1}{2}(E_2^{})^{\mu^{}\nu^{}}dX_{\mu^{}}^{} \wedge dX_{\nu^{}}^{}
\coloneqq\frac{1}{2\sqrt{\gamma (\theta)}}\epsilon^{ij}\frac{\partial X^{\mu^{}}(\theta)}{\partial \theta^{i}}\frac{\partial X^{\nu^{}}(\theta)}{\partial \theta^{j}}dX_{\mu^{}}^{}\wedge dX_{\nu^{}}^{}~.
\label{Nambu bracket}
}
Here, $\theta^i~(i=1,2)$ is an intrinsic coordinate on $M$, 
$\epsilon^{12}=-\epsilon^{21}=1$ is the totally anti-symmetric tensor, and $\gamma(\theta)$ is the determinant of the induced metric on $M$.   
By using the reparametrization degrees of freedom, the component $(E_2^{})^{\mu \nu}(\theta)$ can be also written as~\cite{Iqbal:2021rkn}
\aln{
(E_2^{})^{\mu \nu}(\theta)=t^\mu(\theta) n^\nu(\theta)-n^\mu(\theta) t^\nu(\theta)~,
} 
where $t^\mu(\theta)$ and $n^\mu(\theta)$ are the orthonormal vectors that span the tangent plane at the point $\theta$ on $M$.  
%
%
Note that $\theta_1^{}$ is identified by $\xi$ here, which implies that $t^\mu(\theta)$ matches the normalized tangent vector on the boundary $C$. 
See Ref.~\cite{Iqbal:2021rkn} for more details and also Ref.~\cite{Hidaka:2023gwh} for the generalization to closed $p$-branes.

\subsection{String field theory in the classical continuum limit}  

When analyzing the classical continuum limit, the crucial point is the  interpretation of the kinetic term $(k_0^{}+\hat{H})^{-1}$.     
For sufficiently large $k_0^{}$, it can be expanded as 
\aln{
(k_0^{}+\hat{H})^{-1}=\frac{1}{k_0^{}+1}\left(1+\frac{\hat{H}-1}{k_0^{}+1}\right)^{-1}=\frac{1}{k_0^{}+1}\sum_{n=0}^\infty \left(-\frac{\hat{H}-1}{k_0^{}+1}\right)^n~,
\label{derivative expansion}
}
which is interpreted as a derivative expansion in the continuum limit. 
To illustrate this, it is instructive to see the corresponding example  in ordinary field theory.   
In this case, the translation operator is 
\aln{
\Pi_P^{}\quad \rightarrow \quad \Pi_{\mu}^{\pm}=\exp\left(\pm ia\hat{P}_\mu^{}\right)~,\quad \hat{P}_\mu^{}=-i\frac{\partial}{\partial x^\mu}~.
}
where $\pm $ denotes a direction (orientation) of the translation. 
This leads to 
\aln{
\hat{H}\phi(x)&\coloneqq\frac{1}{2D}\sum_{\mu=1}^D \sum_{s=\pm} \Pi_{\mu}^{i}\phi(x)
\nn
&=\phi(x)+\frac{a}{2D}\sum_{\mu=1}^D\sum_{s=\pm} \mathrm{sgn}(s)\partial_\mu^{}\phi(x)+\frac{a^2}{4D}\sum_{\mu=1}^D\partial_\mu^{2}\phi(x)+{\cal O}(a^3)
\\
&=\phi(x)+\frac{a^2}{4D}\sum_{\mu=1}^D \partial_\mu^{2}\phi(x)+{\cal O}(a^3)~,
}
where the linear term of $a$ vanishes because of the two contributions with opposite signs.  
Thus, we obtain
\aln{
\lim_{a\rightarrow 0}\phi^\dagger(x)\frac{\hat{H}-1}{a^2}\phi(x)=\frac{1}{4D} \phi^\dagger (x)\Box \phi(x)~,\quad \Box=\partial^\mu \partial_\mu^{}~
}
in the continuum limit, and one can see that Eq.~(\ref{derivative expansion}) corresponds to the derivative expansion.   
%
%

\begin{figure}
    \centering
    \includegraphics[scale=0.35]{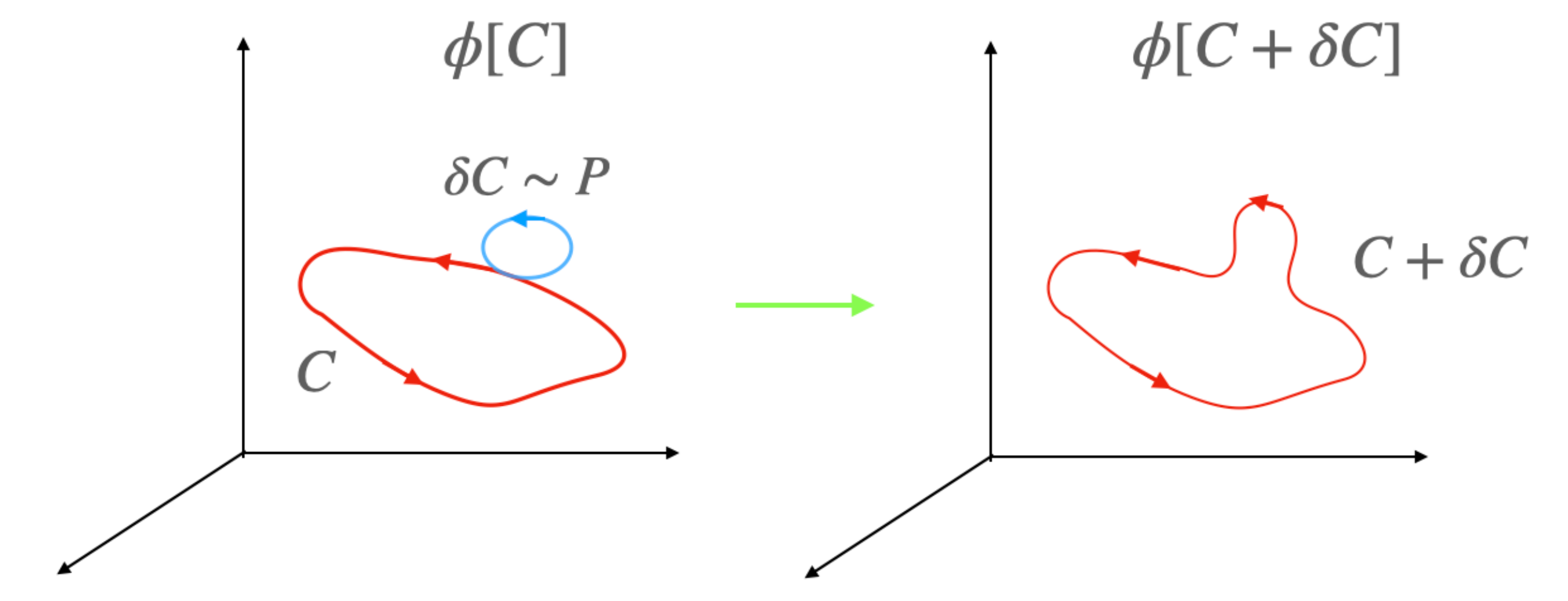}
    \caption{A small change of a loop $C$ by adding an infinitesimal loop (plaquette).  
    The area derivative measures the variation of a functional field $\phi[C]$ under this sort of modifications of loop.   
    }
    \label{fig:area-derivative}
\end{figure}

%
In order to obtain the continuum limit in the string field theory, we have to generalize the notion of  ``derivative" in the space of functional fields $\phi[C]$.    
Actually, such a generalization has been investigated for the loop formulation of non-abelian gauge theory~\cite{Migdal:1983qrz,Makeenko:1980vm,Polyakov:1980ca,Iqbal:2021rkn} and also in string (field) theories~\cite{Kawai:1980qq,Rey:1989ti,Ansoldi:1997cw,Ansoldi:2001km,Aurilia:2002aw}, known as the {\it area derivative}.  
On a cubic lattice, it can be explicitly defined by
\aln{
\frac{\delta \phi[C]}{\delta \sigma^{\mu\nu}(i)}\coloneqq \frac{\phi[C+P_{\mu \nu}(i)]-\phi[C]}{a^2}~,
\label{def of area derivative}
}
for a general loop $C$ (i.e. $C$ can be a self-intersecting loop), where $P_{\mu\nu}^{}(i)$ is a positive oriented plaquette placed in the $\mu$-$\nu$ plane, containing a site $i\in C$. 
We define that $P_{\nu\mu}^{}(i)=P_{\mu\nu}^{-1}(i)$, which corresponds to the anti-symmetric property of the area derivative:
\aln{\frac{\delta}{\delta \sigma^{\mu\nu}(\xi)}=-\frac{\delta}{\delta \sigma^{\nu\mu}(\xi)}~.
}    
One can see that Eq.~(\ref{def of area derivative}) is a natural generalization of the ordinary derivative 
\aln{
\frac{\partial \phi(x)}{\partial x^\mu}\coloneqq \frac{\phi(x+an^\mu)-\phi(x)}{a}
}
to operator acting on a functional field $\phi[C]$ of loops.   
By definition, the area derivative describes the change of a functional $\phi[C]$ when adding a plaquette to the original loop $C$ at around a site $i\in C$.
%
See Fig.~\ref{fig:area-derivative} for a schematic image, where a plaquette (blue) becomes an infinitesimal loop in the continuum limit.  
The important property of the area derivative is~\cite{Iqbal:2021rkn,Hidaka:2023gwh}
\aln{
\frac{\delta}{\delta \sigma^{\mu\nu}(\xi)}\left(\oint_C B_1^{}\right)= (dB_1^{})_{\mu\nu}^{}(X(\xi))
\label{property of area derivative}
}
for $^\forall $ $1$-form field $B_1^{}(X)$, which enables us to calculate the area derivative of an arbitrary function of $\oint_C B_1^{}$ as 
\aln{
\frac{\delta}{\delta \sigma^{\mu\nu}(\xi)}F\left(\oint_C B_1^{}\right)=(dB_1^{})_{\mu\nu}^{}(X(\xi))F'(x)\bigg|_{x=\oint_C B_1^{}}~. 
}
This is analogous to the chain rule of the ordinary derivative $\partial_\mu^{}$.

In the present string field theory~(\ref{loop functional theory}), we have to  consider the fact that the translation operator $\Pi_P^{}$ is nonzero only for plaquettes that share common links with $C$.   
This corresponds to projecting the area derivative~(\ref{def of area derivative}) onto the direction of a shared link as 
\aln{
\frac{\delta}{\delta S^\mu(\xi)}\coloneq t^\alpha(i)\frac{\delta \phi[C]}{\delta \sigma^{\alpha\mu}(i)}~,
\label{projection}
}
where $t^\mu(i)$ is the unit vector at $i\in C$, pointing toward the next site on $C$. 
%
In the continuum limit, $t^\mu(i)$ corresponds to the unit tangent vector $t^\mu(\xi)=\dot{X}^\mu(\xi)/\sqrt{|\dot{X}|^2}$. 

Now, we can evaluate $\hat{H}$ in the continuum limit as
\aln{
\hat{H}\phi[C]&=\frac{1}{2(D-1)|C|}\sum_{P\in C}\Pi_P^{}\phi[C]=\phi[C]+\frac{a^2}{2(D-1)|C|}\sum_{i\in C}\sum_{\mu=1}^D\sum_{s=\pm 1} 
\mathrm{sgn}(s)\frac{\delta \phi[C]}{\delta S^{\mu}(i)}
\nn
&\hspace{4cm}+\frac{(a^2)^2}{4(D-1)|C|}\sum_{i\in C}\sum_{\mu=1}^D\sum_{s=\pm}
\frac{\delta \phi[C]}{\delta S^{\mu}(i)\delta S^{\mu}(i)}+{\cal O}(a^6)
\\
&=\phi[C]+\frac{(a^2)^2}{2(D-1)|C|}\sum_{i\in C}
\sum_{\mu=1}^D\frac{\delta^2 \phi[C]}{\delta S^{\mu}(i)\delta S^{\mu}(i)}+{\cal O}(a^6)~,
}
where the linear term of $a^2$ vanishes because of the two contributions with opposite signs again. 
%
Then, with Einstein notation, we obtain 
\aln{
&\lim_{a\rightarrow 0}\mathrm{Tr}\left(\phi^\dagger[C]\frac{\hat{H}^{}-1}{(a^2)^2}\phi[C]\right)
\nn
=&\frac{1}{2(D-1)|C|}\lim_{a\rightarrow 0}\sum_{i\in C}
\mathrm{Tr}\left(\phi^\dagger [C]\frac{\delta^2 }{\delta S^{\mu}(i)\delta S_{\mu}^{}(i)}\phi[C]\right)
\nn
=&\frac{1}{2(D-1)L[C]}\int_0^{2\pi} d\xi\sqrt{h(\xi)}~\mathrm{Tr}\left(\phi^\dagger [C]\frac{\delta^2}{\delta S^{\mu}(\xi)\delta S_{\mu}^{}(\xi)}\phi[C]\right)~
}
in the continuum limit. 
%
Note that this is reparametrization invariant. 
In addition, the sum over self-avoiding loops becomes the path-integral of  embedding loops as 
\aln{
a^D\sum_{C\in \Gamma}\quad \rightarrow \quad {\cal N}\int {\cal D}X~, 
\label{integral measure}
}
where ${\cal N}$ is an appropriate normalization factor and 
${\cal D}X$ is the ordinary path-integral measure induced by the  diffeomorphism and reparametrization invariant norm
\aln{
||\delta X||^2\coloneq \int_0^{2\pi}d\xi \sqrt{h(\xi)} g_{\mu\nu}^{}(X(\xi))\delta X^{\mu}(\xi) \delta X^{\nu}(\xi)~.
}
We should note that Eq.~(\ref{integral measure}) contains the integral of the center-of-mass coordinates $\{x^\mu\}_{\mu=1}^D$
\aln{\int {\cal D}X=\int d^Dx\sqrt{|g(x)|}\cdots   
} 
because a spacetime translation gives another loop $C'$ for a given loop $C$. 
This is also the reason why we have included the factor $a^D$ in Eq.~(\ref{integral measure}). 

By using the expansion (\ref{derivative expansion}) and performing the field redefinition 
\aln{
\sqrt{\frac{N}{(1+k_0^{})^2(D-1)\beta a^{D-4}}}\phi[C]\rightarrow \phi[C]~,
\label{field redefinition}
}
we finally obtain the continuum theory of Eq.~(\ref{loop functional theory}) as 
\aln{
S_{\rm E}^{\rm SFT}=&V[\phi,\phi^\dagger]-{\cal N}\int {\cal D}X\left[\frac{1}{L[C]}\int_0^{2\pi} d\xi\sqrt{h(\xi)}~\mathrm{Tr}\left(\phi^\dagger[C]\frac{\delta^2}{\delta S^{\mu}(\xi)\delta S_{\mu}^{}(\xi)}\phi[C]\right)
\right]
\nn
&+(\text{higher area-derivative terms})
\label{continuum action}
}
This theory can be regarded as another version of the Landau theory for $1$-form global symmetries proposed in Refs.~\cite{Iqbal:2021rkn,Hidaka:2023gwh,Kawana:2024fsn}.\footnote{
It should also be noted that a similar string field theory was proposed in Ref.~\cite{Kawai:1980qq} for the dual description of the Abelian-Higgs model. 
}
The main difference is the form of kinetic term, and the previously studied  one is  
\aln{
\frac{1}{2L[C]}\int_0^{2\pi} d\xi\sqrt{h(\xi)}\phi^\dagger[C]\frac{\delta^2}{\delta \sigma^{\mu \nu}(\xi)\delta \sigma_{\mu \nu}^{}(\xi)}\phi[C]~,
}
which is simply  constructed by the area derivatives without the projection~(\ref{projection}). 
Naively, this corresponds to the inclusion of self-intersecting loops, i.e. allowing the change of topology of $C$ too.  
%
We believe that this difference in the kinetic terms will become important  when we analyze quantum properties of the model.  

%
The $\mathbb{Z}_N^{}$ $1$-form global symmetry is now expressed as 
\aln{
\phi[C]\quad \rightarrow\quad  \exp\left(\frac{2\pi i}{N}\oint_C \Lambda_1^{} 
\right)\phi[C]~,
\label{one-form transformation}
}
where $\Lambda_1^{}$ is a closed $1$-form with
\aln{
\oint_{C}\Lambda_1^{}\in \mathbb{Z}~. 
}
In fact, one can check that the kinetic term in Eq.~(\ref{continuum action}) is invariant under Eq.~(\ref{one-form transformation}) due to  Eq.~(\ref{property of area derivative}) and the closeness of $\Lambda_1^{}$. 

The potential (\ref{potential expansion}) is also expressed in the continuum limit as 
\aln{
V[\phi,\phi^\dagger]={\cal N}&\int {\cal D}X\left\{\lambda_2^{}[C]\mathrm{Tr}(\phi^\dagger[C]\phi[C])+\lambda_2^{'}[C]\mathrm{Tr}(\phi^\dagger[C])\mathrm{Tr}(\phi[C])+\cdots
\right\}
\nn
+{\cal N}^3\int \left(\prod_{i=1}^3{\cal D}X_i^{}\right)&\delta(C_1^{}-C_2^{}-C_3^{})\bigg\{\lambda_3^{}[\{C_i^{}\}]\mathrm{Tr}(\phi[C_1^{}]^\dagger \phi[C_2^{}]\phi[C_3^{}])+\lambda_3^{'}[\{C_i^{}\}]\mathrm{Tr}(\phi^\dagger [C_1^{}]\phi[C_2^{}])\mathrm{Tr}(\phi[C_3^{}])
\nn
&+\lambda_3^{''}[\{C_i^{}\}]\mathrm{Tr}(\phi^\dagger [C_1^{}])\mathrm{Tr}(\phi[C_2^{}])\mathrm{Tr}(\phi[C_3^{}])\bigg\}+\text{h.c.}+\cdots~,
\label{continuum potential}
}
where we have absorbed additional factors by the field redefinition (\ref{field redefinition}) into the definitions of the coupling constants.  
Note that these coupling constants are dimensionful in general.   

Finally, we can also define a continuum string-field theory on a Lorenzian spacetime $\Sigma_D^{}$ by taking the Wick rotation of Eq.~(\ref{continuum action}) as 
\aln{
S_{\rm SFT}^{}\coloneq iS_{\rm E}^{\rm SFT}=&-V[\phi,\phi^\dagger]+{\cal N}\int {\cal D}X\left[\frac{1}{L[C]}\int_0^{2\pi} d\xi\sqrt{h(\xi)}~\mathrm{Tr}\left(\phi^\dagger[C]\frac{\delta^2}{\delta S^\mu(\xi)\delta S_\mu^{}(\xi)}\phi[C]\right)
\right]
\nn
&+(\text{higher area-derivative terms})~.
\label{Lorentzian action}
}
In Section~\ref{sec4}, we will perform a mean-field analysis based on this Lorentzian action. 
%

\subsection{Mass dimensions}\label{mass dimensions}

As in ordinary quantum field theory, mass dimensions of operators and coupling constants are important for dimensional (qualitative) analysis.   
%
In the following, we represent the mass dimension of a general quantity ${\cal O}$ by $[{\cal O}]$. 
First, note that 
\aln{
\left[\frac{\delta}{\delta S^\mu(\xi)}\right]=\left[\frac{\delta}{\delta \sigma^{\mu\nu}(\xi)}\right]=2~,\quad [{\cal N}{\cal D}X]=-D~
}
by the definitions (\ref{def of area derivative}) and (\ref{integral measure}). 
Then, we find the mass dimension of the string field as
\aln{
[\phi[C]]=\frac{D-4}{2}~. 
\label{mass dimension of string field}
} 
This is consistent with the field redefinition~(\ref{field redefinition}) and also the previous studies~\cite{Iqbal:2021rkn,Hidaka:2023gwh,Kawana:2024fsn}. 
Considering also $[\phi(x)]=\frac{D-2}{2}$ in ordinary field theory, we can guess 
\aln{
[\phi[C_p^{}]]=\frac{D-2(p+1)}{2}
}
for a functional field of closed $p$-brane $C_p^{}$. 
This is indeed the correct mass dimension predicted in the effective field theory of closed $p$-branes in Ref.~\cite{Hidaka:2023gwh}.   

Once the mass dimension of the string field (\ref{mass dimension of string field}) is known, we can also find mass dimensions of various coupling constants. 
For example, in the case of a local potential such as 
\aln{V[\phi,\phi^\dagger]={\cal N}\int {\cal D}X\left\{\mu \mathrm{Tr}(\phi^\dagger \phi)+\lambda \left(\mathrm{Tr}(\phi^\dagger \phi)\right)^2
\right\}~,
}
we have
\aln{[\mu]=4~,\quad [\lambda]=D-8~.
}
In the next section, we will see that $\sqrt{|\mu|}$ corresponds to the world-volume tension of string, and that $[\lambda]=D-8$ could be related to the upper-critical dimension of the system.

\section{Mean Field Analysis}\label{sec4}
In this section, we perform the mean-field analysis of the continuum string field theory~(\ref{Lorentzian action}). 
See also Refs.~\cite{Iqbal:2021rkn,Hidaka:2023gwh,Kawana:2024fsn} for  similar analysis and Ref.~\cite{Yoneya:1980bw} for the large $N$ behaviors. 
%

Let us first clarify the order-parameter of the $\mathbb{Z}_N^{}$ $1$-form global symmetry. 
In the original lattice gauge theory, the order-parameter is the expectation value of the Wilson loop $W[C]$ in the large surface-area limit:
\aln{
\lim_{{\rm Area}(C)\rightarrow \infty}\langle W[C]\rangle ~,
}
which exhibits the Area (Perimeter) law in the unbroken (broken) phase. 
In the dual string field theory, it is replaced by the trace of the string-field $\phi[C]$:
\aln{
\lim_{{\rm Area}(C)\rightarrow \infty}\langle {\rm Tr}(\phi[C])\rangle~. 
}
In fact, by the definition of the source $J[C]^*$ and Eq.~(\ref{loop functional theory}), one can see 
\aln{
\langle W[C]\rangle&=\frac{1}{Z[J]}\frac{\delta Z[J]}{\delta J[C]^*}\bigg|_{J=0}
\nn
&=-\frac{2N}{\beta w[C]}(k_0^{}+\hat{H})^{-1}\langle \mathrm{Tr}(\phi[C])\rangle
\\
&=-\frac{2N}{\beta w[C](k_0^{}+1)}\left(1-\frac{\hat{H}-1}{k_0^{}+1}+\cdots \right)\langle \mathrm{Tr}(\phi[C])\rangle~,
\label{relation of Wilson loop}
}
which implies $\langle W[C]\rangle\propto w[C]^{-1}\langle {\rm Tr}(\phi[C])\rangle$ at the leading-order in the continuum limit.  
The higher-order terms contain the area derivatives, and dimensional analysis indicates that they should be suppressed by a factor of $(a^2\mu )^2$ compared to the leading-order term, where $\mu$ is the (physical) world-volume string tension.  
We will confirm this behavior below. 
%

\subsection{Unbroken phase of $\mathbb{Z}_N^{}$ 1-form symmetry}

Let us consider the unbroken phase of the $\mathbb{Z}_N^{}$ $1$-form global symmetry in the continuum theory (\ref{continuum action}).
Namely, we focus on the parameter space such that the potential $V[\phi,\phi^\dagger]$ has an absolute minimum at $\phi[C]=0$. 
%
In this case, we can show that the equation of motion of Eq.~(\ref{continuum action}) admits a solution that exhibits the Area law in the large surface-area limit~\cite{Iqbal:2021rkn,Hidaka:2023gwh,Kawana:2024fsn}.      
By taking the variation of Eq.~(\ref{continuum action}) with respect to $\phi^\dagger[C]$, we obtain the equation of motion as 
\aln{
\frac{1}{L[C]}\int_{0}^{2\pi} d\xi\sqrt{h(\xi)}\frac{\delta^2 \phi[C]}{\delta S^{\mu^{}}(\xi)\delta S_{\mu^{}}^{}(\xi)}
-\frac{\delta V[\phi,\phi^\dagger]}{\delta \phi^\dagger[C]}=0~.
\label{EOM}
}
For the present purpose, it is sufficient to consider a diagonal configuration\footnote{At mean field level, the off-diagonal property of $\phi[C]$ does not play a significant role in the discussion of the spontaneous breaking of the $\mathbb{Z}_N^{}$ $1$-form global symmetry.  
However, it would become important when addressing quantum aspects of the theory and is left for future investigations.  
}
\aln{
\langle \phi[C]\rangle=\frac{f[C]}{\sqrt{2}}I_N^{}
}
because this configuration preserves the gauge symmetry as guaranteed by the Elitzur's theorem~\cite{Elitzur:1975im}.     
Furthermore, to show the existence of a classical solution exhibiting the Area law, we consider the following ansatz:
\aln{
f[C]=f({\rm Area}(C))~,\quad f(z)\in \mathbb{R}~,
} 
with the boundary condition $f(\infty)=0$.     
Putting this ansatz into Eq.~(\ref{EOM}), we obtain  
\aln{
f''(z)-q(z)f'(z)-\frac{\delta V[f]}{\delta f(z)}=0~,\quad z={\rm Area}(C)~,
\label{EOM in ansatz}
}
where we have used~\cite{Hidaka:2023gwh}
\aln{
\frac{\delta {\rm Area}(C)}{\delta S^{\mu}(\xi)}=t^\alpha(\xi)(E_2^{})_{\alpha^{}\mu^{}}^{}(\xi)\bigg|_{C}^{}=-n_\mu^{}(\xi)~,
}
and  
\aln{
q(z)\coloneqq\frac{1}{L[C]}\int_{0}^{2\pi}d\xi \sqrt{h(\xi)}\frac{\delta n^{\mu}(\xi)}{\delta S^\mu(\xi)}~
}
is a scalar function whose mass dimension is identical to ${\rm Area}(C)^{-1}$. 
By definition, $q(z)$ is determined only by geometrical quantities and does not depend on the coupling constants in the potential.\footnote{As discussed in Refs.~\cite{Iqbal:2021rkn,Hidaka:2023gwh}, we can also construct the effective action for $f(z)$ and obtain the same equation of motion. 
In this case, $q(z)$ can be interpreted as the density of loop configurations whose minimum surface-area is $z$.  
}
Thus, it should behave as $q(z)\rightarrow z^{-1}$ for $z\rightarrow \infty$ on dimensional ground~\cite{Iqbal:2021rkn,Hidaka:2023gwh}.  
Besides, the potential can be approximated as 
\aln{
V[f]\simeq \frac{\mu}{2}{\cal N}\int {\cal D}X f^2 ~,\quad \mu\geq 0
}
for $z\rightarrow \infty$ because of the boundary condition $f(z)\rightarrow 0$. 
Then, Eq.~(\ref{EOM in ansatz}) asymptotically becomes  
\aln{
f''(z)-\mu f(z)\approx 0 \quad \text{for }z\rightarrow \infty~, 
}
which gives the Area law behavior of the string field as 
\aln{
\phi[C]\rightarrow c\times \exp\left(-\sqrt{\mu}\times {\rm Area}(C)\right)\quad \text{for }z\rightarrow \infty~, 
\label{area law behavior}
}
where $c$ is a constant that should be determined by a boundary condition for $z\rightarrow 0$ (small loop limit).  
One can see that $\sqrt{\mu}$ corresponds to the world-volume tension of string.  

A couple of comments are necessary. 
First, when a classical solution exhibits the Area-law behavior as Eq.~(\ref{area law behavior}), the next-leading-order correction in Eq.~(\ref{relation of Wilson loop}) can be estimated as  
\aln{
(\hat{H}-1)\langle \mathrm{Tr}(\phi[C])\rangle\sim & \frac{a^4}{L[C]}\int_0^{2\pi}d\xi \sqrt{h(\xi)}\frac{\delta^2}{\delta S^{\mu}(\xi)\delta S_{\mu}^{}(\xi)}\langle \mathrm{Tr}(\phi[C])\rangle
\nn
\sim & (a^2\mu)^2\langle \mathrm{Tr}(\phi[C])\rangle\quad \text{for } z\rightarrow \infty~,
} 
which is negligible in the continuum limit $a\rightarrow 0$. 
Thus, the Area law of the Wilson loop in the gauge theory is realized as a large surface-area behavior of the classical solution in the dual string field (\ref{Lorentzian action}).  
Second, the above analysis is reliable in the strong coupling regime of the gauge theory, i.e. $g^2\gg 1\leftrightarrow \beta \ll 1$, as suggested by the Hubbard-Stratonovich transformation and also the field redefinition~(\ref{field redefinition});  
After the redefinition, the coupling of a $n$-th order term of $\phi[C]$ in the potential is proportional to $\beta^{\frac{n}{2}}\propto g^{-n}$, and we can rely on the perturbative calculations for $g\gg 1$.  
In other words, the strong-coupling limit in the gauge theory corresponds to the weak-coupling limit in the dual string field theory. 
In this sense, the realization of the Area law is quite natural (and maybe trivial) because it is well-known that the Wilson loop in the lattice gauge theory exhibits the Area law by the strong coupling expansion, as reviewed in Appendix~\ref{strong coupling}. 

For $g^2\ll 1\leftrightarrow \beta \gg 1$, on the other hand, the mean-field analysis based on the classical action Eq.~(\ref{Lorentzian action}) would be no longer reliable, and quantum corrections must be taken intro account.      
%

\subsection{Broken phase of $\mathbb{Z}_N^{}$ 1-form symmetry}\label{broken phase}

When the potential $V[\phi,\phi^\dagger]$ has nontrivial minima, the string field $\phi[C]$ develops a nonzero VEV $\langle \phi[C]\rangle \neq 0$, leading to the spontaneous breaking of the $\mathbb{Z}_N^{}$ $1$-form global symmetry. 
In general, the $\mathbb{Z}_N^{}$ symmetry implies the existence of $N$ degenerate (gauge-invariant) minima 
\aln{
\phi[C]=\frac{v}{\sqrt{2}}e^{\frac{2\pi i n}{N}}I_N^{}
~,\quad v\in \mathbb{R}~,\quad \quad n=1,2,\cdots,N~. 
\label{degenerate VEV}
}
As in ordinary quantum field theory, our primary interest is to identify low-energy effective degrees of freedom, and to find an effective theory of it. 
In the present string field theory, a candidate for such a low-energy mode is the phase modulation given by  
\aln{
\phi[C]=\frac{v}{\sqrt{2}}\exp\left(i\oint_C A_1^{}\right)I_N^{}~,\quad A_1^{}\coloneq A_\mu^{}(X)dX^\mu ~.
\label{phase modulation}
}  
The effective theory of $A_1^{}$ must respect the following $1$-form shift symmetry 
\aln{
A_1^{}\quad \rightarrow \quad A_1^{}+\frac{1}{N}\Lambda_1^{}~,\quad d\Lambda_1^{}=0~,  
\label{shift symmetry}
}
with $\oint_C \Lambda_1^{}\in 2\pi \mathbb{Z}$ because this transformation  corresponds to the change of the VEV by a $\mathbb{Z}_N^{}$ factor as in  Eq.~(\ref{degenerate VEV}). 

By putting Eq.~(\ref{phase modulation}) into Eq.~(\ref{Lorentzian action}), we obtain the effective action as
\aln{
S_{\rm eff}[A_1^{}]=-\frac{v^2}{2}\int_{\Sigma_D^{}}F_2^{}\wedge \star F_2^{}-\int_{\Sigma_D^{}}U\left(A_1^{}\right)\star 1
\label{effective action of A}
}
where $F_2^{}=dA_1^{}$ and $U(A_1^{})$ is an effective potential with the periodicity    
\aln{
U\left(A_1^{}+\frac{\Lambda_1^{}}{N}\right)=U(A_1^{})~.
}
%
See Appendix~\ref{calculation of A action} for the detail derivation.    
%
%
%
In the low-energy limit, we can further approximate the effective potential by using the Villain formula~\cite{Villain:1977} and introducing  an auxiliary $1$-form field $f_1^{}$ and $(D-2)$-form field $B_{D-2}^{}$ as   
\aln{
&\exp\left(-i\int_{\Sigma_D^{}} U(A_1^{})\star 1\right)
\nn
\approx &\int {\cal D}f_1^{}\int {\cal D}B_{D-2}^{}\exp\left(-i\int_{\Sigma_D^{}}\frac{\lambda_2^{}}{2}\left(A_1^{}-\frac{f_1^{}}{N}\right)\wedge \star \left(A_1^{}-\frac{f_1^{}}{N}\right)-\frac{i}{2\pi}\int_{\Sigma_D^{}}B_{D-2}^{}\wedge df_1^{}\right)~,
\label{Villain}
}
where $\lambda_2^{}$ is a coupling constant. 
See again Appendix~\ref{calculation of A action} for the details. 
In this form, we can eliminate $f_1^{}$ by using the equation of motion as
\aln{
\frac{f_1^{}}{N}=A_1^{}+\frac{N(-1)^{D-2}}{2\pi \lambda_2^{}}\star dB_{D-2}^{}~,
} 
which leads to the dualized form of the effective action:
\aln{
S_{\rm eff}^{}[A_1^{},B_{D-2}^{}]=&-\frac{v^2}{2}\int_{\Sigma_D^{}}F_{2}^{}\wedge \star F_{2}^{}
\nn
+\int_{\Sigma_D^{}}&\left[\frac{N^2(-1)^{D-2}}{8\pi^2 \lambda_2^{}}dB_{D-2}^{}\wedge \star dB_{D-2}^{}-\frac{N^2(-1)^{D-2}}{4\pi^2 \lambda_2^{}}d(B_{D-2}\wedge \star dB_{D-2}^{})-\frac{N}{2\pi}B_{D-2}\wedge dA_1^{}
\right]~.
}
The (topological) property of ground state is described by neglecting all the kinetic terms, and we have 
\aln{
S_{\rm eff}^{}[A_1^{},B_{D-2}^{}]\approx -\frac{N}{2\pi}
\int_{\Sigma_D^{}}B_{D-2}\wedge dA_1^{}~.
\label{topological field theory}
}
which is nothing but a $\mathrm{BF}$-type topological field theory. 
In addition to the $\mathbb{Z}_N^{}$ $1$-form global symmetry (\ref{shift symmetry}), this topological field theory has an emergent $\mathbb{Z}_N^{}$ $(D-2)$-form global symmetry:
\aln{
B_{D-2}^{}\quad \rightarrow \quad B_{D-2}^{}+\frac{\Lambda_{D-2}^{}}{N}~,\quad d\Lambda_{D-2}^{}=0 
\label{emergent symmetry}
}
with 
\aln{\int_{C_{D-2}}^{}\Lambda_{D-2}^{}\in 2\pi \mathbb{Z}~, 
}
where $C_{D-2}^{}$ is a $(D-2)$-dimensional closed subspace in $\Sigma_D^{}$. 
The charged objects (symmetry operators) are the Wilson-loop and -surface respectively: 
\aln{
W[C]=\exp\left(i\oint_C A_1^{}\right)~,\quad V[B_{D-2}^{}]=\exp\left(i\int_{C_{D-2}^{}} B_{D-2}^{}\right)~. 
}
The topological theory (\ref{topological field theory}) exhibits topological order depending on the topology of spacial manifold $\Sigma_{D-1}^{}$. 
See Refs.~\cite{Gomes:2023ahz,Bhardwaj:2023kri,Hidaka:2023gwh,Kawana:2024fsn} and references therein for more details of topological order.  
%

\subsection{Topological defects}\label{defect}
The emergence of the $\mathbb{Z}_N^{}$ $(D-2)$-form symmetry (\ref{emergent symmetry}) is deeply related to the existence of topological defects/excitations in the broken phase.   
In the lattice gauge theory (\ref{lattice action}), it is known that there exist codimension $2$ topological configurations as local minima of the lattice action, i.e. the center vortex~\cite{Greensite:2003bk,Yoneya:1978dt}. 
A center-vortex configuration can be also regarded as symmetry operator (i.e. Gukov-Witten operator) of the $\mathbb{Z}_N^{}$ $(D-2)$-form global symmetry;   
When a center vortex links with a Wilson loop $W[C]$, it produces a $\mathbb{Z}_N^{}$ transformation $W[C]\rightarrow e^{\frac{2\pi i}{N}n}W[C],~n\in \mathbb{Z}$.   
Duality between the lattice gauge theory and string field theory indicates  that there should exist similar topological configurations in the latter theory. 
In the following, we explicitly construct a topologically nontrivial static configuration in the continuum string-field theory (\ref{continuum action}).   
In Appendix~\ref{center vortex}, we also review the construction of center-vortex solution in the lattice gauge theory.

\begin{figure}
    \centering
    \includegraphics[scale=0.3]{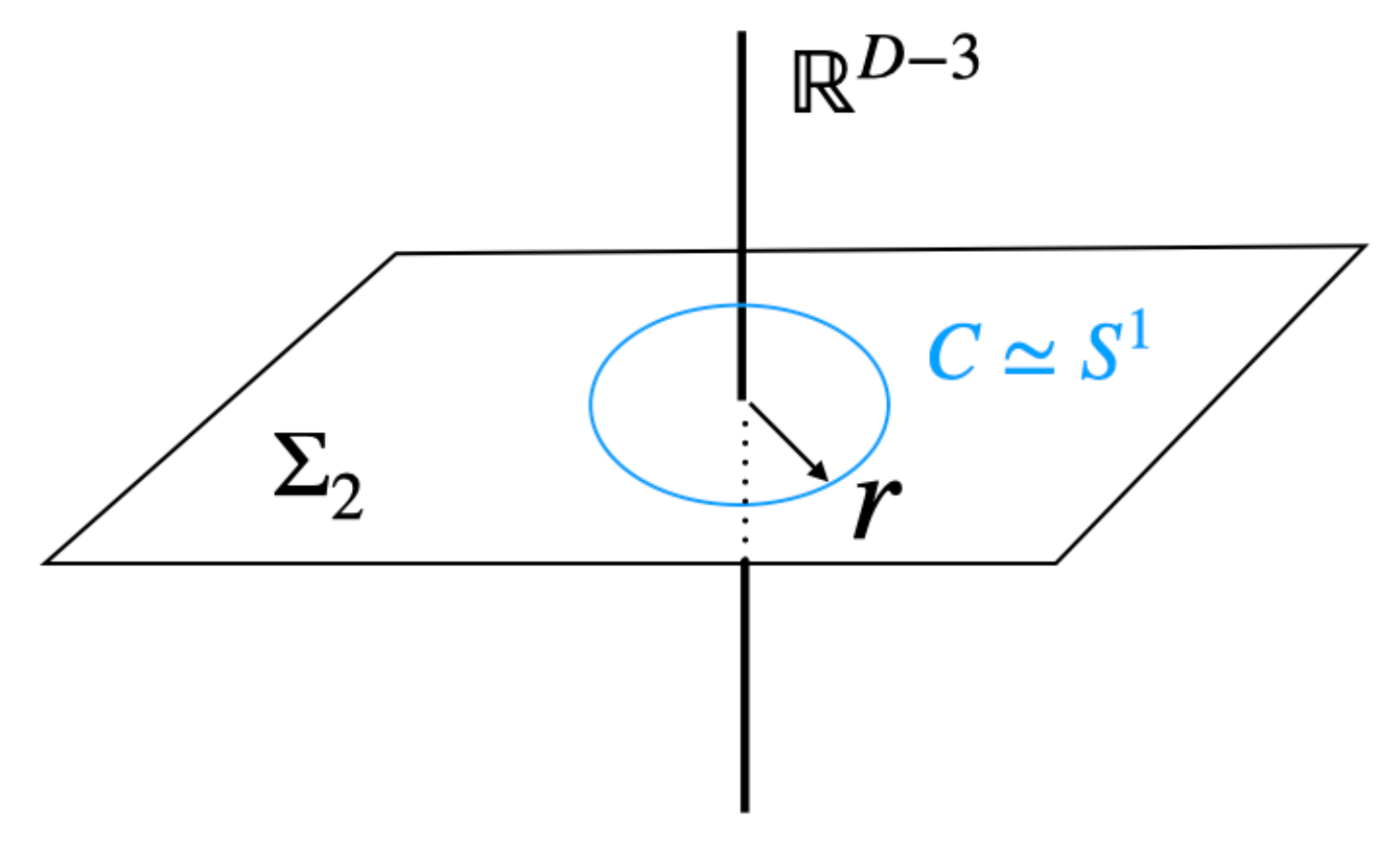}
     \includegraphics[scale=0.3]{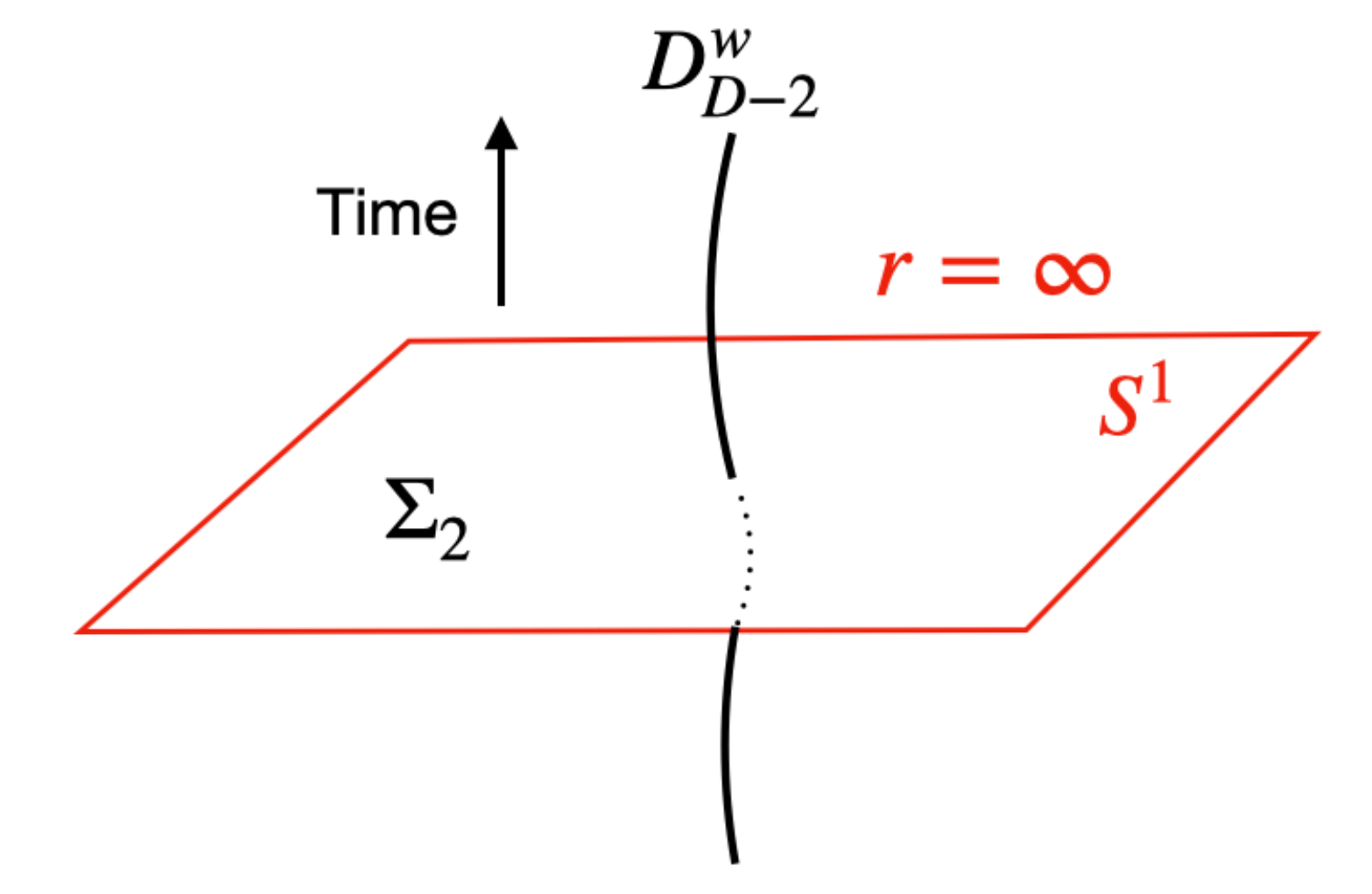}
    \caption{Left: A topologically non-trivial static configuration in the continuum string field theory (\ref{continuum action}) in a time-slice.  
    The blue circle corresponds to the embedded loop $C^{}\simeq S^1$ characterized by the radius $r$.  
    Right: The linking between the world-volume $D^W_{D-2}$ of the topological defect and $S^{1}$. 
    }
    \label{fig:soliton}
\end{figure}

For simplicity, we consider the flat spacetime $\Sigma_D^{}=\mathbb{R}\times \mathbb{R}^{D-1}$ and represent the space as 
\aln{
\mathbb{R}^{D-1}=\mathbb{R}^{D-3}\times S^{1}\times [0,\infty)~,
}
where $r\in [0,\infty)$ denotes the radius of $S^1$.  
See the left panel in Fig.~\ref{fig:soliton} for example. 
In particular, we represent the $2$-dimensional flat subspace as 
\aln{
\Sigma_{2}^{}\coloneqq S^{1}\times [0,\infty)~,
}
which has a boundary $\partial \Sigma_{2}^{}= \Sigma_{2}^{}|_{r=\infty}^{}=S^{1}$ at space infinity.  
%

Then, we can consider a loop $C^{}$ embedded as $r=$fixed with an intrinsic coordinate $\xi\in [0,2\pi)$ as shown by a blue circle in Fig.~\ref{fig:soliton}.  
In this case, a loop is specified only by the radius $r$ and it corresponds to the reduction of the path-integral measure as 
\aln{
{\cal N}\int {\cal D}X \quad \rightarrow \quad {\cal N}\int_0^\infty dr r^{}
~,
\label{minispace}
} 
We then consider a following ansatz:
\aln{
\phi[C_p^{}]=\frac{v}{\sqrt{2}}\exp\left(i\int_{S^1}A_1^{W}\right)I_N^{}
\coloneqq \frac{v}{\sqrt{2}}\exp\left(if(r)\int_{0}^{2\pi}d\xi\right)I_N^{}=\frac{v}{\sqrt{2}}\exp\left(2\pi if(r)\right)I_N^{}~,
\label{static configuration}
}
with boundary conditions
\aln{
f(0)=0~,\quad f(\infty)=\frac{n}{N}~,\quad n\in \mathbb{Z}~. 
}
Namely, Eq.~(\ref{static configuration}) is a topological configuration which interpolates one minimum $\phi[C]=v/\sqrt{2}$ to another $ve^{\frac{2\pi i}{N}n}/\sqrt{2}$.  
The corresponding topological charge is given by
\aln{
Q_{D-2}^{}=\frac{N}{2\pi}\int_{\Sigma_2^{}}dA_1^{W}=\frac{N}{2\pi}\oint_{S^1}A_1^{W}\bigg|_{r=\infty}^{}=n\in \mathbb{Z}~,
}
where we have used the Stokes theorem in the second equality.

Denoting the world-volume of the defect as 
\aln{
D^W_{D-2}\coloneq \mathbb{R}\times \mathbb{R}^{D-3}~,
}
we can also relate $A_1^{W}$ to the world-volume manifold as 
\aln{
\delta_{2}^{}(D^W_{D-2})=\frac{N}{2\pi Q_{D-2}^{}}dA_1^W~,\label{world-volume form}
}
where $\delta_{2}^{}(D^W_{D-2})$ is the Poincare-dual form defined by
\aln{
\int_{D^W_{D-2}} f_{D-2}^{}=\int_{\Sigma_D^{}}f_{D-2}^{}\wedge \delta_{2}^{}(D^W_{D-2})~
}
for $^\forall~(D-2)$-form $f_{D-2}^{}$. 
In particular, we can check 
\aln{
\int_{\Sigma_{D}^{}}\delta_{2}^{}(D^W_{D-2})\wedge \delta_{D-2}^{}(\Sigma_{2}^{})=\int_{\Sigma_{2}}\delta_{2}^{}(D^W_{D-2})=\frac{N}{2\pi Q_{D-2}^{}}\int_{\Sigma_{2}}dA^W_{1}=1~,
}
which represents the linking between $D_{D-2}^W$ and $S_{}^{1}$, as depicted  in the right panel in Fig.~\ref{fig:soliton}.  

The area derivative of the ansatz~(\ref{static configuration}) is evaluated as 
\aln{
\frac{\delta \phi[C_p^{}]}{\delta S^\mu(\xi)}=
it^\alpha (dA_1^W)_{\alpha \mu}\phi[C_p^{}]=-i\frac{1}{r}\frac{df}{dr}\delta_\mu^{r}\phi[C_p^{}]~,
}
which leads to the effective action of $f(r)$ as 
\aln{
S_{\rm SFT}^{}[f]
&=-{\cal N}\int_0^\infty drr\left[\frac{v^2}{2r^2}\left(\frac{df}{dr}\right)^2+U(f)\right]
\\
&=-{\cal N}v^4\int_0^\infty dyy\left[\frac{1}{2y^2}\left(\frac{df}{dy}\right)^2+\overline{U}(f)\right]~,
}
where we have introduced a dimensionless radius $y=vr$ and potential $\overline{U}\coloneqq v^{-6}U$. 
The dimensionless equation of motion is now given by 
\aln{
\frac{1}{y}\frac{d}{dy}\left(\frac{1}{y}\frac{df}{dy}\right)-\frac{\partial \overline{U}}{\partial y}=0~.
\label{defect equation}
}
In general, the periodic potential $\overline{U}(f)=\overline{U}(f+m/N)$ is given by a Fourier series of $\cos(2\pi N k f)$ and $\sin(2\pi Nk f)$ with $k\in \mathbb{N}$.  
For a concrete example, let us consider the simplest case  
\aln{
\overline{U}(f)=\overline{U}_A^{}(f)\coloneq \overline{\lambda}\left(1-\cos\left(2\pi N f\right)\right)~,
\label{cosine potential}
} 
where $\overline{\lambda}$ is a dimensionless parameter. 
One can check that this potential is produced by the interaction terms listed in Eq.~(\ref{determinant terms}) under the ansatz Eq.~(\ref{static configuration}). 
\begin{figure}
    \centering
    \includegraphics[scale=0.65]{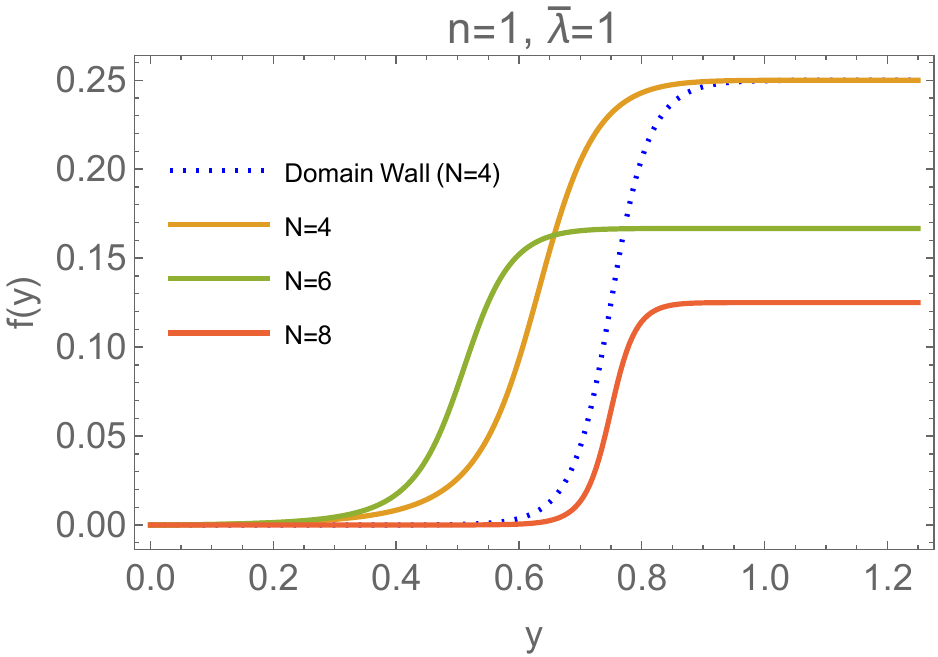}
    \caption{
    Field profiles of the topological defect in the continuum string field theory.  
    }
    \label{fig:profile}
\end{figure}
\noindent 
In Fig.~\ref{fig:profile}, we show the numerical plots of $f(y)$ for $N=4$ (blue), $6$ (orange), and $8$ (green) with $n=1$ and $\overline{\lambda}=1$.      
For comparison, we also present the domain-wall solution (dotted blue) given by 
\aln{
\frac{d^2f}{dy^2}-\frac{\partial \overline{U}_A^{}}{\partial y}=0~,\quad f(0)=0~,\quad f(\infty)=\frac{1}{N}~
}   
with $N=4$. 
While these field profiles essentially have the same functional behaviors,  we should note that the way they divide the space is quite different: 
In the usual domain-wall case, it spreads over $\mathbb{R}^{D-2}$ in space, whereas the above topological defect spreads over $\mathbb{R}^{D-3}$ subspace like a vortex, as shown in the left figure in Fig.~\ref{fig:soliton}.   

We can similarly obtain the low-energy effective theory in the presence of the topological defect. 
Instead of Eq.~(\ref{phase modulation}), the phase modulation $A_1^{}$ is introduced by 
\aln{\phi[C]=\frac{v}{\sqrt{2}}\exp\left(i\oint_C (A_1^{}+A_1^W)\right)I_N^{}~, 
} 
and the low-energy effective theory can be obtained by replacing $A_1^{}$ in Eq.~(\ref{topological field theory}) by  $A_1^{}+A_1^W$ as 
\aln{S_{\rm eff}^{}[A_1^{},B_{D-2}^{}]\approx \frac{N}{2\pi}\int_{\Sigma_{D}^{}}B_{D-2}^{}\wedge dA_1^{}+Q_{D-2}^{}\int_{D_{D-2}^{W}}B_{D-2}^{}~,
}
where we have used Eq.~(\ref{world-volume form}). 
The second term represents the interaction between $B_{D-2}^{}$ and the topological defect. 
%

\subsection{Critical dimensions}

We comment on critical dimensions of the present theory.~\footnote{
In Ref.~\cite{Iqbal:2021rkn}, the upper critical dimension was estimated by  dimensional analysis, and it is claimed to be consistent with the Parisi's heuristic argument.  
Here, we provide more details of this argument, along with the estimations of lower critical dimension.  
}
In Ref.~\cite{Parisi:1978mn}, the lower and upper critical dimensions of (pure) non-abelian gauge theories are heuristically estimated as 
\aln{
D^{\rm L}_c=4~,\quad \text{and}\quad D^{\rm U}_c=8~
}
by employing the concept of Hausdorff dimension. 
The argument in Ref.~\cite{Parisi:1978mn} is based on the viewpoint that the lattice gauge theory can be also interpreted as a theory of random surfaces, 
%
and we can define the Hausdorff dimension of random surfaces as follows: 
A crucial observation is the large-step behavior of the average surface area $\langle S\rangle_n^{}$ of $n$-step triangulations (plaquettes), and it grows as  
\aln{
\langle S\rangle_n^{}\coloneq L(n)^2\sim a^2 n^{1/2}\quad \text{for $n\rightarrow \infty$} 
\label{length and steps}
}
as long as the central-limit theorem holds.  
Here, $L(n)$ denotes the effective size of $n$-step random surfaces. 
When we consider Hausdorff dimension, we inversely interpret Eq.~(\ref{length and steps}) as 
\aln{
n(L)\sim \left(\frac{L}{a}\right)^4 \quad \text{for $L\rightarrow \infty$}~, 
}
which means that the number of triangulations increases as the fourth power of the size $L$, leading to the Hausdorff dimension of random surfaces as  
\aln{
D_{\rm H}^{\mathrm{RS}}=4~.
}
Then, it was naively argued in Ref.~\cite{Parisi:1978mn} that the Hausdorff dimension would also correspond to the lower critical dimension $D_c^{\rm L}=D_H^{}=4$ by a simple reasoning that it is the case in scalar field theory with continuous symmetries, i.e. $D_c^{\mathrm{L}}=D_H^{\mathrm{RW}}=2$. (Here, $\mathrm{RW}$ means ``random walk".)   
%
%
%
However, the free-energy argument discussed below 
 suggests $D_c^{\mathrm{L}}=2$, which seems to be more reliable because it corresponds to a discrete-symmetry version of the Coleman-Mermin-Wagner theorem~\cite{Gaiotto:2014kfa,Lake:2018dqm}.

The estimation of upper critical dimension is a bit more non-trivial.  
It is based on the consideration of a dimension of $A\cap B$ for general two objects $A$ and $B$: 
\aln{
d(A\cap B)=d(A)+d(B)-D~. 
} 
For example, for $D=3$ and $d(A)=d(B)=2$, we have $d(A\cap B)=2+2-3=1$, i.e. 
the intersection of two surfaces in $3$-dimensional space is a line.   
By applying this formula to two random surfaces $\mathrm{RS}_A^{}$ and $\mathrm{RS}_B^{}$, we obtain 
\aln{
d(\mathrm{RS}_A^{}\cap \mathrm{RS}_B^{})=2D_{\rm H}^{\mathrm{RS}}-D=8-D~,
}
which implies that for $D\geq 8$ there is no intersection of random surfaces typically, and the critical behaviors would be well described by the  mean-field dynamics around the Gaussian fixed point, as we have studied  in the previous sections.      
For $D<8$, however, two random surfaces have an intersection in general, indicating that interactions between surfaces become crucial for the critical behaviors of the system.      
From these heuristic observations, the upper critical dimension of random surfaces is estimated as $D_c^{\rm U}=8$ in Ref.~\cite{Parisi:1978mn}.

We can also estimate critical dimensions based on free energy and renormalization group arguments    
%
in the continuum string field theory:    
As we have seen in the previous section, the present theory allows a $(D-2)$-dimensional topological object in the broken phase, which looks like a particle excitation when $D=2$.   
By denoting the mass of such an excitation by $\Delta m$, the change of the free energy can be estimated as 
\aln{
\Delta F\approx \Delta m-\beta^{-1}\log\left(\frac{V_2^{}}{a^2}\right)~,
} 
where the second term represents the entropy change and $V_2^{}$ is the volume of the system. 
This means $\Delta F<0$ in the thermodynamic limit $V_2^{}\rightarrow \infty$, and leads to the proliferation of topological excitations. 
Thus, we can conclude 
\aln{D_c^{\rm L}=2\quad (\text{by the free energy argument})~.
}
The same argument can be applied to a general discrete $p$-form global symmetry and we have $D_c^{\rm L}=p+1$~\cite{Lake:2018dqm}.

The estimation of the upper critical dimension is less rigorous.   
In Section~\ref{mass dimensions}, we have seen that the mass dimension of coupling constants of the local quartic interactions is $[\lambda]=D-8$ by dimensional analysis. 
Thus, if we were able to neglect all the other non-local interactions in Eq.~(\ref{continuum potential}), a naive renormalization group argument  predicts $D_c^{\rm U}=8$, analogous to $D_c^{\mathrm{U}}=4$ in the ordinary $\phi^4$ theory.
Interestingly, this agrees with the estimation by the analysis of Hausdorff dimension above. 
However, as already mentioned in Ref.~\cite{Iqbal:2021rkn}, this estimation  is too naive and the existence of non-local interactions provably alters the critical behaviors of the system. 
Further investigations are left for future publications. 

\section{Summary and Discussion}\label{sec5}

We have studied the classical continuum limit of the string field theory dual to the $\mathrm{SU}(N)$ lattice gauge theory and performed the mean-field analysis. 
The resultant continuum theory is analogous to the string field theory for $1$-form global symmetries studied in  Ref.~\cite{Iqbal:2021rkn,Hidaka:2023gwh,Kawana:2024fsn}. 
The key idea is the utilization of the area derivative, which is a natural generalization of the ordinary derivative $\partial/\partial x^\mu$, and we have found that the kinetic term in the continuum theory is expressed  by the d'Alembert operator constructed by the area derivatives. 
Then, we have seen that the continuum theory retains the $\mathbb{Z}_N^{}$ $1$-form global symmetry of the gauge theory due to the unique property of the area derivative. 
In the unbroken (broken) phase of the $\mathbb{Z}_N^{}$ $1$-symmetry, we have explicitly shown that the classical string field exhibits the Area (Perimeter) law, which implies that the confinement/deconfinement  transition is described in terms of the spontaneous breaking of the $\mathbb{Z}_N^{}$ $1$-form symmetry. 
Furthermore, we have explicitly constructed a codimension $2$ topological defect in the broken phase and shown that the effective theory is given by a $\mathrm{BF}$-type topological field theory coupled to the topological defect. 
The effective theory has an emergent $\mathbb{Z}_N^{}$ $(D-2)$-form global symmetry and can exhibit topological order.

There are many critical issues to be addressed:  
We know little about quantum natures of the dual string field theory.  
%
%
In particular, even the calculation of the tree-level propagator seems to be nontrivial and we hope that the results from the first-quantization theory~\cite{Ansoldi:2001km,Ansoldi:1995hp,Aurilia:2002aw} would be useful for this purpose.   
Besides, the check of unitarity is a crucial issue for the consistency of the present theory.  
 We can naively expect that unitarity is preserved in the continuum string field theory because it is directly constructed by the (lattice) gauge theory, but a rigorous proof requires more elaborated studies as in other string field theories.  
It is also interesting to see whether we can develop a Feynman-diagram technique for the perturbative calculations in the present string field theory. 

Second, the impact of non-local interactions on the critical behaviors is not fully investigated even at mean-field level.    
Since they generally contain odd-order terms of $\phi[C]$, the phase transition would become first-order, as is suggested by many numerical studies of lattice gauge theories. 
It is interesting and important to study if we can also find another critical points in the presence of these nonlocal interactions and investigate  universality class in the (mean) string field theory.  
 
Finally, there is also a variety of extensions of the present framework: 
Generalization to $p$-form global symmetries is a natural direction; For a given system with a $p$-form global symmetry, it is interesting to see if we can construct a similar dual field theory.   
We expect that such a dual field theory would be described by the $p$-brane field theory constructed in Ref.~\cite{Hidaka:2023gwh} in the low-energy limit.  
Furthermore, another promising direction is the study of systems without Lorentz invariance.
In such systems, it is known that there is no one-to-one correspondence between the generators of broken symmetry and Nambu-Goldstone modes~\cite{Nielsen:1975hm,Schafer:2001bq,Miransky:2001tw,Nambu:2004yia,Watanabe:2011ec}. 
The complete relation can be understood by considering the expectation value of the commutation relation of broken generators, and this has been extended to the case with $p$-form global symmetry~\cite{Hidaka:2020ucc}. 
It is interesting to study how low-energy effective theory is derived
from the perspective of field theory of branes.

\section*{Acknowledgements}
We thank Yoshimasa Hidaka and Tamiaki Yoneya for the fruitful discussions and comments. 
This work is supported by KIAS Individual Grants, Grant No. 090901.   
%

\appendix 

\section{Strong coupling expansion}\label{strong coupling}
Here, we review the strong coupling expansion in the lattice gauge theory. 
Our focus is the expectation value of the Wilson loop 
\aln{\langle W[C]\rangle=\frac{1}{Z}\int [dU]W[C]e^{-S_\mathrm{E}^{}[U]}~.
\label{expectation of the Wilson loop}
}
In the strong coupling limit $g^2\gg 1\leftrightarrow \beta \ll1$, we expand the lattice partition function as
\aln{\exp\left(-\frac{\beta}{2N}\sum_{P:\text{positive}} (W[P]+W^\dagger [P])
\right)=\sum_{n=0}^\infty\frac{1}{n!}\left(-\frac{\beta}{2N}\sum_{P:\text{positive}} (W[P]+W^\dagger [P])\right)^n~,
\label{strong coupling expansion}
}
and calculate the multiple group integrals.  
The key properties are 
\aln{\int dUU_{ij}^{}=0~,\quad \int dU U^\dagger_{ij}U_{kl}^{}=\frac{1}{N}\delta_{jk}^{}\delta_{il}~. 
\label{group integral}
}
From the first equation, one sees that if a link variable $U_{\hat{\mu}}^{}(i)$ appears just once in the integrand (\ref{expectation of the Wilson loop}) with the expansion (\ref{strong coupling expansion}), it gives zero contribution. 
This means that each of the link variables $U_{\hat{\mu}}(i)$ must have its companion $U_{\hat{\mu}}^\dagger (i)$ in order to have a non-vanishing contribution in Eq.~(\ref{expectation of the Wilson loop}).  
For a given Wilson loop $W[C]$, this means that non-zero contributions appear  when a set of plaquettes (with the same orientation) constitute a surface $S$ enclosed by $C$.  
The corresponding $\beta$ dependence is given by
\aln{\left(\frac{\beta}{2N}\right)^{\#\text{ of plaquettes}}~.
}
Then, each paired links give $1/N$ contributions due to the second equation in Eq.~(\ref{group integral}), which totally gives 
\aln{\left(\frac{1}{N}\right)^{\#\text{ of links}}~.
}
After the group integrals, we are left with the trace of the color indices $\sum_{i=1}^{N}1=N$ for every site on the surface $S$, which produces 
\aln{
N^{\#\text{ of sites}}~.
} 
As a result, a surface $S$ with $\partial S=C$ gives the contribution to the Wilson loop as
\aln{
\langle W[C]\rangle=\left(\frac{\beta}{2N}\right)^{\#\text{ of plaquettes}}\left(\frac{1}{N}\right)^{\#\text{ of links}}
N^{\#\text{ of sites}}~.
} 
Apparently, the leading order contribution comes from the minimal surface of  $C$ which has 
\aln{\#\text{ of plaquettes}&=\frac{LT}{a^2}~,
\\
\#\text{ of links}&=\frac{(L+a)T}{a^2}+\frac{(T+a)L}{a^2}~,
\\
\#\text{ of sites}&=\frac{(L+a)(T+a)}{a^2}~,
}
which lead to
\aln{\langle W[C]\rangle=N\left(\frac{\beta}{2N^2}\right)^{\frac{LT}{a^2}}=Ne^{-\sigma{\rm Area}(C)}~,
}
where
\aln{\sigma=-\frac{1}{a^2}\log \left(\frac{\beta}{2N^2}\right)~.
}
This is the derivation of the Area law in the strong coupling limit. 
However, we should note that this does not imply the confinement in the continuum limit because the latter is achieved for $g\rightarrow 0$ along with $a\rightarrow 0$. 
It is known that the $1/g^2$ expansion has a finite convergence radius and we cannot take the continuum limit actually.

\section{Calculations of the effective action}
\label{calculation of A action}
Here we provide the detail calculations of the effective action of $A_1^{}$. 
In the following, we consider the flat spacetime $g_{\mu\nu}^{}=\eta_{\mu\nu}^{}$ for simplicity. 

Let us first focus on the kinetic term. 
By putting Eq.~(\ref{phase modulation}) into the kinetic term in the continuum action~(\ref{Lorentzian action}), we have
\aln{
-\frac{v^2}{2}{\cal N}\int {\cal D}X\frac{1}{L[C]}\int_0^{2\pi} \sqrt{h(\xi)}t^\alpha(\xi)t^\beta(\xi) {F_{\alpha}^{}}^\mu  (X(\xi)) F_{\beta \mu}^{}(X(\xi))~,
\label{kinetic term of A}
} 
where $F_{2}^{}=dA_1^{}=\frac{1}{2}F_{\mu\nu}^{}dX^\mu\wedge dX^{\nu}$.  
By introducing the center-of-mass coordinates and relative coordinates as
\aln{
x^\mu &\coloneqq\frac{1}{L[C]}\int_{0}^{2\pi} d\xi\sqrt{h}X^\mu(\xi)~,  
\\
Y^\mu(\xi)&\coloneqq X^\mu(\xi)-x^\mu~,
}
we have 
\aln{
{F_{\alpha}^{}}^{\mu}(X)F_{\beta \mu}^{}(X)=\int \frac{d^Dk}{(2\pi)^D}e^{ik\cdot (x+Y)}
{\tilde{F}_{\alpha}^{*}}{}^{\mu}(k)\tilde{F}_{\beta \mu}^{}(k)~,
}
where $\tilde{F}_{\alpha\beta}^{}(k)$ is the Fourier mode. 
By putting this into Eq.~(\ref{kinetic term of A}), we obtain
\aln{
&\int d^Dk\left(\int \frac{d^Dx}{(2\pi)^D}~e^{ik\cdot x}\right){\tilde{F}_{\alpha}}{}^{\mu}(k)\tilde{F}_{\beta \mu}^{}(k)
{\cal N}\int {\cal D}Y\frac{1}{L[C]}\int_{0}^{2\pi} d\xi\sqrt{h}t^{\alpha}(\xi)t^{\beta}(\xi)e^{ik\cdot Y(\xi)}
\nn
=&{\tilde{F}_{\alpha}}{}^{\mu}(0)\tilde{F}_{\beta \mu}^{}(0)\frac{{\cal N}}{L[C]}\int_{0}^{2\pi} d\xi\sqrt{h}\langle t^\alpha(\xi)t^\beta(\xi)\rangle~,
}
where
\aln{
\langle t^\alpha(\xi)t^\beta(\xi)\rangle &=\int {\cal D}Y t^\alpha (\xi)t^\beta(\xi)
\nn
&=\lim_{\varepsilon\rightarrow 0+}\int {\cal D}Y e^{-\varepsilon L[C]} \frac{dY^\alpha}{ds}\frac{dY^\beta}{ds}\quad (s=\text{proper length})
\\
&=\lim_{\varepsilon\rightarrow 0+}\lim_{t\rightarrow s}\frac{\partial}{\partial s}\frac{\partial }{\partial t}G^{\alpha\beta}(s-t;\varepsilon)
\label{world line path integral}
}
Here, we have introduced a weight factor $e^{-\varepsilon L[C]}$ to make the path-integral well-defined for large loop configurations, and 
\aln{
G^{\alpha \beta }(s-t)=\int {\cal D}Y e^{-\varepsilon L[C]}Y^\alpha(s)Y^\beta(t)
} 
is the $2$-point correlation function.  
For the present discussion, we do not need to calculate this exactly. 
What we need is the property 
$G^{\alpha \beta}(s;\varepsilon)=\eta^{\alpha \beta}G(s;\varepsilon)$, which holds as long as spacetime symmetry is unbroken.    
A scalar function $G(s;\varepsilon)$ (and its derivatives) can generally contain UV divergences~\cite{Iqbal:2021rkn}, but we can always absorb (or renormalize) it by the normalization factor ${\cal N}$ as
\aln{
{\cal N}\frac{1}{L[C]}\int_{0}^{2\pi} d\xi\sqrt{h}\lim_{\varepsilon\rightarrow 0+}\lim_{t\rightarrow s}\frac{\partial}{\partial s}\frac{\partial }{\partial t}G(s-t;\varepsilon)=1~,
}
which leads to the kinetic term of $A_1^{}$ as
\aln{
-\frac{v^2}{2}\tilde{F}^{\mu\nu}(0)\tilde{F}_{\mu \nu}^{}(0)=-\frac{v^2}{2}\int_{\Sigma_D^{}}F_2^{}\wedge \star F_2^{}~.
}
This is nothing but the Maxwell theory.

Next, let us consider the potential of $A_1^{}$.  
As a concrete example, we focus on the $N$-th order terms listed in Eq.~(\ref{determinant terms}).  
By putting Eq.~(\ref{phase modulation}) into these terms, we have (up to overall normalization)
\aln{
\int {\cal D}X\cos\left(N\oint_C A_1^{}\right)=\int d^Dx U(A_1^{})
}
where
\aln{
U(A_1^{})=\int {\cal D}Y~\cos\left(\oint_C A_1^{}(x+Y)\right)~.
}
%
One can check that this is invariant under the $\mathbb{Z}_N^{}$ shift~(\ref{shift symmetry}).  
In the low-energy limit, we can use the Villain formula~\cite{Villain:1977} as 
\aln{&\exp\left(-i\int d^Dx U(A_1^{})\right)\approx \sum_{n\in \mathbb{Z}}\exp\left(-i{\cal N}\int {\cal D}X\left\{1-\frac{1}{2}\left(N\oint_C A_1^{}-2\pi n\right)^2\right\}
\right)
}
which can be rewritten by a path-integral form by introducing two auxiliary higher-form fields as 
\aln{
=&\int {\cal D}f_1^{}\int {\cal D}B_{D-2}^{}\exp\left(-i{\cal N}\int {\cal D}X\left\{1-\frac{1}{2}\left(\oint_C (NA_1^{}-f_1^{})\right)^2\right\}-\frac{i}{2\pi}\int_{\Sigma_D^{}}B_{D-2}^{}\wedge df_1^{} 
\right)~
}
with $\int_{C}f_1^{}\in 2\pi \mathbb{Z}$. 
Then, by repeating the similar calculations to the kinetic term, we obtain
\aln{
=&\int {\cal D}f_1^{}\int {\cal D}B_{D-2}^{}\exp\left(-i\int_{\Sigma_D^{}}\left[1-\frac{\lambda_2^{}}{2}\left(A_1^{}-\frac{f_1^{}}{N}\right)\wedge \star \left(A_1^{}-\frac{f_1^{}}{N}\right)\right]\star 1-\frac{i}{2\pi}\int_{\Sigma_D^{}}B_{D-2}^{}\wedge df_1^{} 
\right)~,
} 
where $\lambda_2^{}$ is a coupling constant.

\section{Center Vortex}\label{center vortex}
We review the center-vortex solution in the lattice gauge theory. 
See also Ref.~\cite{Greensite:2003bk} and references therein for more details. 
By taking the variation of the lattice action~(\ref{lattice action}) with respect to $A_\mu^{a}(i)$, we obtain the equation of motion as 
\aln{
\sum_{\nu\neq \mu}^D\mathrm{Tr}\left[T^a\left(U_{\mu\nu}(i)-U_{\mu\nu}^\dagger(i)\right)
\right]=0~,
}
which allows a very simple but interesting class of solutions given by 
\aln{
U_{\hat{\alpha}}^{}(i)=g(i)Z_{\hat{\alpha}}^{}(i)g^\dagger(i+\hat{\alpha})
\label{lattice solution}
}
with
\aln{
Z_{\hat{\alpha}}^{}(i)=\exp\left(\frac{2\pi i}{N}n_{\hat{\alpha}}^{}(i)\right)\in \mathbb{Z}_N^{}~,\quad n_{\hat{\alpha}}^{}(i)\in \mathbb{Z}~,
}
where $g(i)\in \mathrm{SU}(N)$. 
However, not all the solutions are local minima of the action. 
We can see it by considering fluctuations 
\aln{U_{\hat{\mu}}^{}(i)=Z_{\hat{\mu}}^{}(i)V_{\hat{\mu}}^{}(i)~,\quad V_{\hat{\mu}}^{}(i)=\exp\left(i\sum_{a}A_\mu^{a}(i)T^a\right)~,
}
which leads to the lattice action 
\aln{S_\mathrm{E}^{}&=\beta \sum_P\left[1-\frac{1}{2N}\left(Z_P^{}V[P]+Z_P^{*}V[P]^*\right)\right]
\\
&=\beta \sum_P\left[1-\frac{1}{2N}\left(Z_P^{}+Z_P^{*}
\right){\rm Tr}\left(I_N^{}-\frac{1}{2}F^{\mu\nu}F_{\mu\nu}\right)+{\cal O}(F^4)
\right]
\\
&=\beta \sum_P\left[1-\cos\left(\frac{2\pi}{N}n_P^{}\right)+\cos\left(\frac{2\pi}{N}n_P^{}\right){\rm Tr}\left(F^{\mu\nu}F_{\mu\nu}\right)+{\cal O}(F^4)
\right]~,
}
where 
\aln{Z_P^{}=\exp\left(\frac{2\pi i}{N}n_P^{}\right)~,\quad n_P^{}=n_{\hat{\mu}}^{}(i)+n_{\hat{\nu}}^{}(i+\hat{\mu})+n_{-\hat{\mu}}^{}(i+\hat{\mu}+\hat{\nu})+n_{-\hat{\nu}}^{}(i+\hat{\nu})~.
}
One can see that the solution (\ref{lattice solution}) is a local minimum of the action when $\cos\left(\frac{2\pi}{N}n_P^{}\right)>0$ for $^\forall P$. 
For a non-vortex plaquette, i.e. $n_P^{}=0$, this is trivially satisfied. 
\begin{figure}
    \centering
    \includegraphics[scale=0.3]{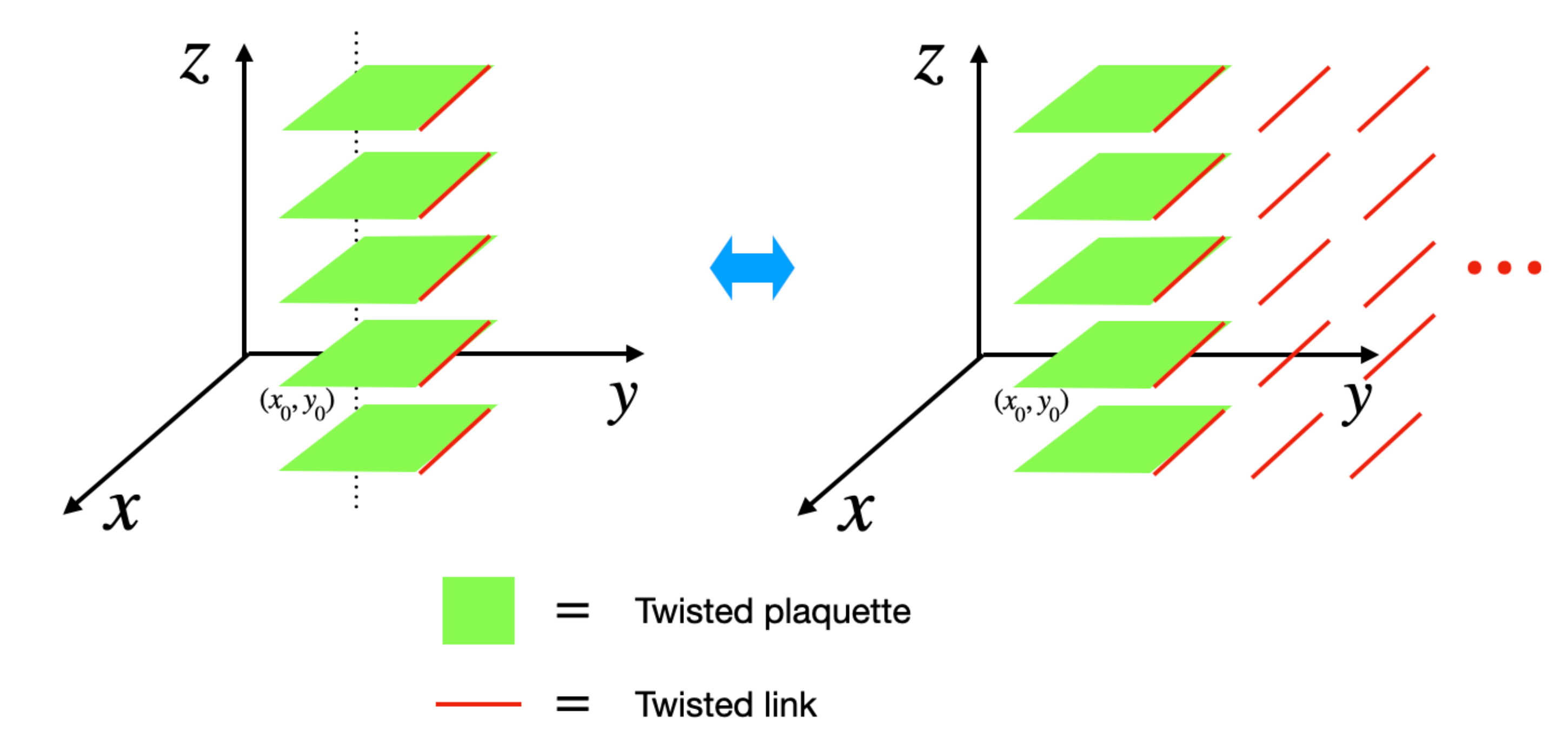}
    \caption{A center vortex solution in a $3$-dimensional cubic lattice. 
    }
    \label{fig:center vortex}
\end{figure}
On the other hand, we need 
\aln{
\cos\left(\frac{2\pi}{N}n_P^{}\right)>0\quad \leftrightarrow \quad \frac{n_p^{}}{N}<\frac{1}{4}\quad \text{or}\quad \frac{N-n_p^{}}{N}<\frac{1}{4}~
} 
for a vortex plaquette with $n_P^{}\neq 1$. 
This condition cannot be satisfied for $N\leq 4$, while several nonzero values of $n_P^{}$ are allowed for $N\geq 5$, e.g. $n_P^{}=1,4$ for $N=5$. 
A simplest center vortex-solution in $D=4$ can be obtained by piling up twisted plaquettes $U[P]\rightarrow Z_P^{}U[P]$  with $n_P^{}=1$ toward the $z$-direction at a point $(x_0^{},y_0^{})$ in the $x$-$y$ plane. 
This is shown in the left panel in Fig.~\ref{fig:center vortex}, where the red link represents a twisted link for a given twisted plaquette (green).  
Since this configuration is a $1$-dimensional object in space, it corresponds to a $2$-dimensional object in spacetime. 
Note also that the left configuration in  Fig.~\ref{fig:center vortex} is equivalent to the right configuration in Fig.~\ref{fig:center vortex}, where all the links with $y=y_0^{}$ and $x> x_0^{}$ are twisted too, because such additional twists cancel each other in all the plaquette for $x>x_0^{}$. 

\bibliographystyle{TitleAndArxiv}
\bibliography{Bibliography}

\end{document}